\pdfoutput=1
\documentclass[a4paper,12pt]{article}
\usepackage{jheppub}

\usepackage[english]{babel}
\usepackage{amsmath}   
\usepackage{graphicx}  
\usepackage{dcolumn}   
\usepackage{bm}        
\usepackage{amssymb}   
\usepackage{epsfig}
\usepackage{epstopdf}  
\usepackage{subfigure}
\usepackage[utf8]{inputenc}
\usepackage{mdframed}
\usepackage{amsmath}
\usepackage{wasysym}
\usepackage{tabu}
\usepackage[toc]{appendix}
\usepackage{color,soul} 
\usepackage{datetime}
\usepackage{IEEEtrantools}
\usepackage{tikz}
\usetikzlibrary{calc, 3d}

\newcommand{\dif}{\mathop{}\!\mathrm{d}}

\title{Imaging Higher Dimensional Black Objects} 
\author{Thomas Hertog,} 
\author{Tom Lemmens and}
\author{Bert Vercnocke}
\emailAdd{firstname.lastname@kuleuven.be}
\affiliation{Institute for Theoretical Physics, KU Leuven,
Celestijnenlaan 200D, B-3001 Leuven, Belgium}
\date{\today}

\abstract{
We develop a systematic ray-tracing method which can be used to explore a wide class of higher-dimensional multi-center black objects through the shape of their shadows. As a proof of principle, we test our method by imaging black holes and black rings in five dimensions. Two-dimensional slices of the three-dimensional shadows of those five-dimensional black objects not only show new phenomena, such as the first image of a toroidal horizon, but also offer a new viewpoint on black holes and binary systems in four dimensions.
}

\begin{document}

\setcounter{tocdepth}{2}

\maketitle

\section{Introduction}\label{sec:intro}

It is well known that the black hole uniqueness theorems in General Relativity (GR) are specific to vacuum GR in four dimensions. In more than four dimensions or when gravity is coupled to matter fields, a plethora of new black hole solutions comes into play. In more than four spacetime dimensions, black objects exist with distinct horizon topologies but with the same masses and spins \cite{Emparan:2008eg}. Coupling GR to matter fields gives rise to black hole solutions with scalar, fermionic or vector hair; configurations of matter fields outside the horizon\footnote{See e.g. \cite{Herdeiro:2015aa,Volkov:2016aa} for recent reviews.}. In this context, one even has compact objects without horizons from self-gravitating bosonic configurations known as boson stars. Other extensions of GR, involving higher derivatives, quantum effects, etc, similarly lead to modifications of the unique GR horizon. This richness of compact objects in extensions of GR has spawned vast fields of study, ranging from exploring the intrinsic theoretical properties of those solutions to fleshing out all the many possibilities for testing GR with recent and future observations of black holes in nature.

String theory combines all of the above richness. Its extra dimensions and the many matter fields arising in compactifications mean that the wide variety of gravitational black hole like solutions is already evident at the classical level. The low-energy limit of string theory, supergravity, has a vast set of solutions that are absent in 3+1 vacuum GR.
To leading order in the string length expansion, supergravity theories are two-derivative Lagrangians with couplings to matter fields in up to eleven spacetime dimensions. 
The solution space contains the three types of solutions mentioned above: non-spherical horizon topologies, hairy black holes, and horizonless solutions. The causal structure and  other spacetime properties of the vast number of solutions has remained largely unexplored. The reason is simply the complexity of most of those solutions. Even the earliest known black ring in pure GR in 4+1 dimensions \cite{Emparan:2001wn} has a non-integrable geodesic problem \cite{Grunau:2012ai}. Many more of the solutions with and without horizons have a high degree of asymmetry with at best one (timelike or null) Killing vector, but in general no other isometries.

With this paper, we aim to start a systematic numeric exploration of the many black hole like solutions descending from string theory, by investigating their null geodesic structure and using this to image the compact objects. We develop and present a backwards ray tracing code and a visualization scheme adapted to axisymmetric but otherwise arbitrary asymptotically flat, supersymmetric solutions of  five-dimensional ungauged supergravity coupled to vector multiplets, described in \cite{Gutowski:2004yv,Elvang:2004ds,Gauntlett:2004qy,Bena:2004de} and reviewed in \cite{Bena:2007kg}. 

We choose this class of five-dimensional supersymmetric solutions for two reasons. First, five is the lowest number of dimensions allowing both new horizon topologies and smooth horizonless solutions. Second, due to supersymmetry the solution space is linear in the sense that one can add sources freely and thus identify solutions with an arbitrary number of centers. Whereas specific solutions of pure GR in higher dimensions need to be run case by case, our code enables the imaging of many solutions in a fast manner: The restriction to axial symmetry, or more properly one timelike and two rotational Killing vectors in 5D, is enough to reduce the numerical integration time to the order of hours on a normal laptop, for an image of $1000\times1000$ pixels. 

To demonstrate our method, we produce ray-traced images of Breck\-enridge-Meyers-Peet-Vafa (BMPV) black holes \cite{Breckenridge:1996is} and supersymetric black rings of Emparan, Elvang, Mateos and Reall \cite{Elvang:2004rt}. To the best of our knowledge, these are the first ray-traced images of higher-dimensional objects\footnote{Analytic result on the black hole shadow exist for Myers-Perry black holes \cite{Papnoi:2014aaa}.} and the first images of black rings altogether.  
As a verification of our method, we compare the numerically generated BMPV shadows with analytically computed shadows cast by BMPV black holes and find perfect agreement. Two-dimensional slices of these ray-traced images exhibit all known features of rotating black holes in four dimensions, including the characteristic shape of near-extremal images due to frame dragging. 

Our images of black rings reveal new features. Depending on which two-dimen\-sional slicing one considers, the images can serve different purposes. While the black ring horizon has topology $S^1 \times S^2$, one possible 2D slicing has the topology of two disconnected two-spheres. Imaging this reveals the same key features known for binary systems in four dimensions, such as disconnected shadows with characteristic `eyebrows' \cite{Nitta:2011in,Yumoto:2012kz,Bohn:2014xxa,Patil:2016oav,Shipley:2016omi,Wang:2017qhh,Cunha:2018gql,Cunha:2018cof}. This opens up a new analytic avenue for exploring four-dimensional black hole binaries. Other 2D slicings of black ring shadows show the first examples of shadows of toroidal horizons\footnote{A toroidal shadow of a horizon with spherical topology has been observed earlier, see \cite{Cunha:2016bjh}.}, which can appear as concentric annuli in the two-dimensional projection. We do not compare to analytic results for black rings as those are only available for a restricted set of geodesics \cite{Grunau:2012ri}.

The rest of this paper is organized as follows. In section \ref{sec:vis}, we give the details of our visualization scheme and integration method. In section \ref{sec:susyBH}, we discuss the BMPV black hole metric, its ray-traced images and compare to analytic results.  In section \ref{sec:susyBR} we discuss our images of supersymmetric black rings from different viewing angles and explain the salient features. We end with a summary and outlook on the many future imaging paths forward in section \ref{sec:conclusions}. The appendices contain the following supplementary material: Appendix \ref{sec:metrics} has a technical review of the full class of multi-center solutions to explain our conventions. We provide a derivation of the geodesic equations for axisymmetric multi-center solutions in appendix \ref{sec:geodesics}, and we end with the calculation of the geodesic equations and the shadow contour for BMPV black holes in appendix \ref{sec:BMPV_exact}.

\section{Integration and Visualization Procedure}\label{sec:vis}

\begin{figure}[hbt]
	
	\begin{center}
		\includegraphics[scale=0.25]{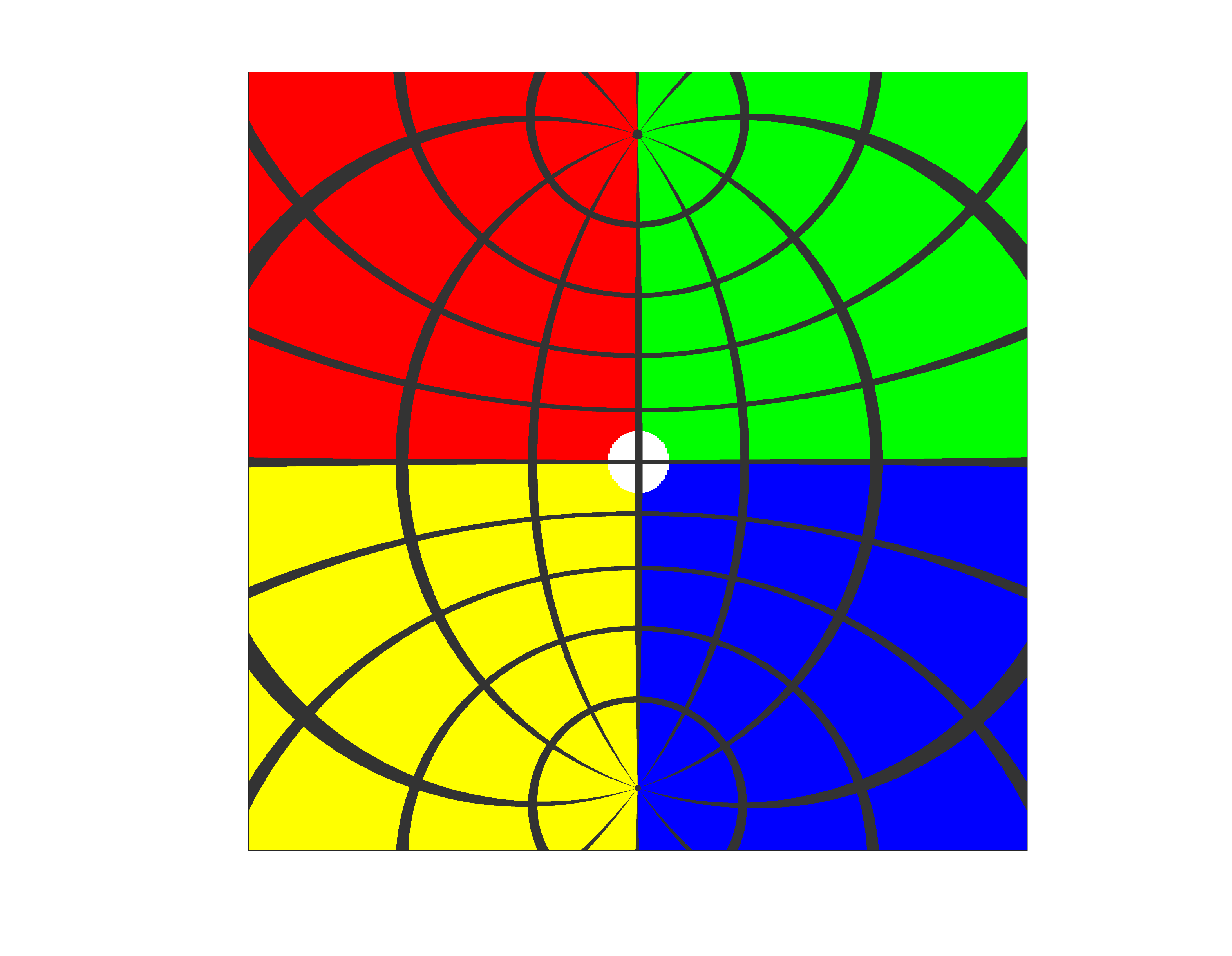}
	\end{center}
	
	\caption{The view of an asymptotic observer if there is no central object present. This view was generated by using the numerical algorithm on five-dimensional Minkowski space as a consistency check. The location of the colors have been chosen to match the colors of \cite{Cunha:2015yba,Cunha:2018acu}. }
	\label{fig:empty}
	
\end{figure}

To create an image we use an extension of the setup discussed in \cite{Bohn:2014xxa,Cunha:2015yba,Cunha:2018acu}.  Readers who want to get to the physics can skip this section and just keep in mind that following the procedure with no central object gives the reference image in figure \ref{fig:empty}.

The goal is to assign a color to each pixel on a 2D slice of the three-dimensional screen constructed in the local sky of an asymptotic observer. 
We do this in three steps. First we associate a geodesic to each direction in the observer's local sky (section \ref{sec:tech}). Next we follow this geodesic numerically until it either escapes or falls behind a horizon (section \ref{ssec:integration}). And finally we associate a color to this geodesic depending on where it ended up (section \ref{ssec:color}). 

\subsection{The Local Sky}\label{sec:tech}

\begin{figure}[hbt]
	\begin{center}
		\resizebox {0.75\linewidth} {!} {
			\begin{tikzpicture}
			
			\def\L{2}
			\def\D{7.5}
			\def\ang{30}
			\def\l{1}
			
			\newcommand{\axesFour}[1]{
				\draw [->] (0,0) -- (#1,0) node[above] {$x_{1}$};
				\draw [->] (0,0) -- (0,#1) node[above] {$x_{3}$};
				\draw [->] (0,0) -- (45:0.75*#1) node[above] {$x_{2}$};
				\draw[->] (0,0) -- (135:0.75*#1) node[above] {$x_{4}$};
			}
			
			\axesFour{\L}

			\draw (0,0) -- (\ang:\D);
			
			\begin{scope}[shift={(\ang:\D)}]
			\begin{scope}[rotate=\ang]
			\draw [->] (0,0) -- (-\l,0) node[above] {$x$};
			\draw [->] (0,0) -- (0,\l) node[above] {$z$};
			\end{scope}
			\draw [->] (0,0) -- (-135:0.75*\l) node[below] {$y$};
			\draw[->] (0,0) -- (-45:0.75*\l) node[above] {$w$};
			
			\end{scope}

			\end{tikzpicture}}
	\end{center}
	
	\caption{Geometry of the two different Cartesian coordinate axes. The $(x_{1},x_{2},x_{3},x_{4})$ form a global Cartesian coordinate system and the $(x,y,z,w)$ are a set of Cartesian axes located at our distant observer.}
	\label{axes}
	
\end{figure}
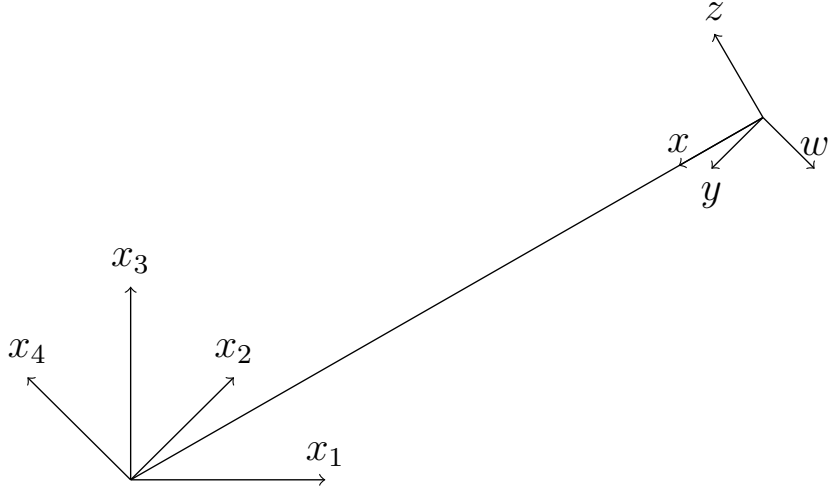

The first step is to associate a geodesic to each direction in the local sky of our observer in four-dimensional space. To bring this into practice, we will relate a set of local Cartesian coordinates for the asymptotic observer $(x,y,z,w)$, to an appropriate global Cartesian coordinate system $(x_1,x_2,x_3,x_4)$, see figure \ref{axes}. We then express the asymptotic direction (a vector in the observer's frame) to null geodesics in the global spacetime coordinates.

The local set of Cartesian axes $(x,y,z,w)$ has its origin at the position of the observer. Without loss of generality, we can choose this to be 
\begin{equation}
 (x^1,x^2,x^3,x^4)_{\rm observer} = (d\sin i,0,d\cos i,0)\,,
\end{equation}
with the inclination $i$ and the Cartesian distance $d$ from the origin. A direction in which the observer is looking is then represented by a unit vector in this local Cartesian frame. Before we can continue we need to fix these local Cartesian axes in some way. We can begin by putting the $x$ axis in the radial direction towards the origin of the spacetime. Due to our choice of position this lies in the $(x_{1},x_{3})$ plane of the global Cartesian coordinates. We can pick the $z$ axis to be in the $(x_{1},x_{3})$ plane perpendicular to the $x$ axis, and we orient it such that it points in the positive $x_{3}$ direction. We take the $y$--axis and $w$--axis to be aligned with the $-x_{2}$--axis and $-x_{4}$--axis respectively. This construction is illustrated in figure \ref{axes}.

Next we represent a vector in this local frame in the spacetime coordinates. If we start with a vector in the local frame with components $(v_{x},v_{y},v_{z},v_{w})$ we can calculate what its components are in the global Cartesian frame with some simple geometry:
\begin{IEEEeqnarray}{rCl}
	v_{x_{1}} &=& -v_{x}\sin i - v_{z}\cos i\nonumber\\
	v_{x_{2}} &=& -v_{y} \nonumber\\
	v_{x_{3}} &=& -v_{x}\cos i + v_{z}\sin i\nonumber\\
	v_{x_{4}} &=& -v_{w}.\label{eq:velocities}
\end{IEEEeqnarray}

At this point we have a map from a local vector to a null geodesic in the spacetime. To get a flat image we map a set of rectilinear coordinates $(a,b,c) \in [0,1]^3$ to a local vector. To do this we use a Mercator like projection similar to the one used in \cite{Bohn:2014xxa,Cunha:2015yba,Cunha:2018acu}:
\begin{IEEEeqnarray}{rCl}
	v_{x} &=& k_{0} \label{eq:cami}\\
	v_{y} &=& k_{0}(2a-1)\tan\left(\frac{\alpha_{y}}{2}\right) = k_{0} \bar{v}_{y}\\
	v_{z} &=& k_{0}(2b-1)\tan\left(\frac{\alpha_{z}}{2}\right) = k_{0} \bar{v}_{z}\\
	v_{w} &=& k_{0}(2c-1)\tan\left(\frac{\alpha_{w}}{2}\right) = k_{0} \bar{v}_{w}, \label{eq:camf}
\end{IEEEeqnarray}
the $\alpha_{y}$, $\alpha_{z}$ and $\alpha_{w}$ effectively represent different camera apertures and the $k_{0}$ is a normalization constant which we will use to fix the energy of the geodesic. We can now form an image made up of discrete voxels by discretizing the three coordinates $(a,b,c)$. This is useful to get an idea of the shape of the shadow which a full 5D observer would see. But this is not a very convenient way to study the lensing effects since different areas will be hidden behind and inside each other. To study the lensing effects it is better to take 2D slices of the full 3D $(a,b,c)$ space and then discretize this 2D subspace to get an image made out of pixels.

The multi-center solutions of our interest are naturally written in Gibbons-Hawking coordinates $(r,\theta,\phi,\psi)$. To apply the above steps, we bring the asymptotic form of the metric into standard Cartesian coordinates on flat space by the coordinate redefinition:
\begin{IEEEeqnarray}{rCl}
	x_{1} &=& 2\sqrt{r}\cos\left(\phi + \frac{\psi}{2}\right)\cos\left(\frac{\theta}{2}\right),\nonumber\\
	x_{2} &=& 2\sqrt{r}\sin\left(\phi + \frac{\psi}{2}\right)\cos\left(\frac{\theta}{2}\right),\nonumber\\
	x_{3} &=& 2\sqrt{r}\cos\left(\frac{\psi}{2}\right)\sin\left(\frac{\theta}{2}\right),\nonumber\\
	x_{4} &=& 2\sqrt{r}\sin\left(\frac{\psi}{2}\right)\sin\left(\frac{\theta}{2}\right).\label{eq:globalcoords}
\end{IEEEeqnarray}
The coordinate distance and inclination are then:
\begin{equation}
 r_{\rm observer} = \frac{d^2}4,\qquad \theta_{\rm observer}=-2i+\pi\label{eq:def_observer_position_inclination}.
\end{equation}
In appendix \ref{sec:geodesics} we show how we to set up the initial value problem at fixed energy $E$ in Gibbons-Hawking coordinates. The result is that we solve the system of equations given by \eqref{eq:geod} starting from the initial conditions \eqref{eq:init}.

\subsection{Numerical Integration}\label{ssec:integration}

We now have set up an initial value problem (IVP) for each triplet $(a,b,c)$ in camera space. To solve the IVP we use the following procedure:
\begin{enumerate}
	\item We integrate the IVP by using for example the RKDP method as implemented in the ODE45 function of the MATLAB ODE suite (in the future a more specialized integrator can be used).
	\item To handle coordinate singularities we use event location to stop the integration when we get close and decrease the tolerance of the integrator. We estimate when we have moved far enough away from the singularity and check if we have indeed done so when we reach that point. If we have, we increase the tolerance back to the original value and continue, otherwise we estimate again and keep the tolerance.
	\item We end the integration when we reach one of two possible exit conditions. The first is if the geodesic is back at the same radial coordinate at which it started, this signifies that it has escaped to infinity. The second condition is if we are within a chosen small Cartesian distance from a horizon, in this case we say that the geodesic has fallen behind the horizon.
	\item To estimate the integration error we calculate the RMS of $\xi(\tau)-1$ along each geodesic (see appendix \ref{sec:geodesics}). If this is larger than a predetermined tolerance we reject this geodesic and restart the integration with more stringent tolerances.  
\end{enumerate}
Those four steps provide a map from the camera coordinates $(a,b,c)$ to either the horizon or to some position far away from the object.

\subsection{Color Coding}\label{ssec:color}

As a third  and final step, we associate the final position of each geodesic to a color. All the geodesics which fell behind the horizon are marked black. The geodesics which escaped from the central object are colored according to the given reference image. In principle this image is three-dimensional. We will always restrict to a 2D subspace first and apply the colour coding afterwards as follows. First we convert the 4D final positions of the escaped geodesics to Cartesian coordinates. Next we take the 3D subspace containing the radial direction from our observer and the two directions corresponding to  the chosen hypersurface. In this 3D subspace we can go to standard spherical coordinates and map the polar angle to the vertical direction of the image and the azimuthal direction to the horizontal direction of the image. This gives us an RGB value for each escaped geodesic. Using this visualization an observer would see figure \ref{fig:empty} if there is no central object in the way.

\section{Supersymmetric Black Holes}\label{sec:susyBH}
Perhaps the simplest non-trivial metric in the multi-center class we describe is the Breckenridge-Myers-Peet-Vafa (BMPV) black hole \cite{Breckenridge:1996is}. It describes the 5D version of the extremal Reissner-Nordstrom metric.

\subsection{Metric}
We work with the form of the metric adapted to the multi-center extension described in appendix \ref{sec:metrics}: 
\begin{equation}
 ds^2 = - Z^{-2/3} (dt + k)^2 + Z^{-1/3} ds^2_4 (\mathbb{R}^4),
\end{equation}
with the functions 
\begin{equation}
 Z = 1 + \frac {Q}{4r}\,,\qquad k = \frac{J}{8r} (d\psi + (1 + \cos \theta) d\phi)\,.
\end{equation}
Four-dimensional flat space is written as a particular Gibbons-Hawking fibration \cite{Hawking:1976jb,Gibbons:1979zt} over $\mathbb{R}^3$ with radial coordinate $r$:
\begin{IEEEeqnarray}{rCl}
 ds^2_4 (\mathbb{R}^4)&=& r (d\psi + (1+\cos \theta) d\phi)^2+ r^{-1} \left(dr^2 + r^2 
 (d\theta^2 +\sin^2 \theta d\phi^2)\right).\label{eq:GH_R4}
\end{IEEEeqnarray}
The relation to Cartesian coordinates $x^1,x^2,x^3,x^4$ is given by \eqref{eq:globalcoords}.

\subsection{Physical Properties}

The BMPV black hole has an electric charge $Q$ and two angular momenta. We label those as $J_{12}$ and $J_{34}$ in standard Euclidean coordinates $x^1,x^2,x^3,x^4$. Supersymmetry imposes the extremality constraint on the mass and equality of the angular momenta:
\begin{equation}
 M = \frac{\pi}{4 G_5}Q\,, \qquad  J_{12} = J_{34}\equiv \frac{\pi}{4 G_5}J\,,
\end{equation}
with $G_5$ the 5D Newton constant. We will chose units such that  $\frac{\pi}{4 G_5} =1$.  The fact that the solution has rotation might seem surprising from GR intuition, since in 3+1 dimensions, black holes can have only one independent rotation that is required to vanish by invariance under supersymmetry. In 4+1 dimensions however, black holes can have two independent rotations in two orthogonal planes; supersymmetry requires only one combination of them to vanish. 

The horizon has the topology of $S^3$, a three-sphere, with horizon area $A_H = 4 \pi \sqrt{Q^3 - J^2}$.
As for the Kerr metric, too large values of the angular momenta can lead to pathologies. Positive horizon area requires $J \leq Q^{3/2}$. For larger angular momenta, the solution becomes `over-spinning' and has naked Closed Timelike Curves. Those can be resolved invoking higher-dimensional effects \cite{Gibbons:1999uv,Herdeiro:2000ap,Herdeiro:2002ft,Drukker:2004zm}. We choose to focus on imaging BMPV black holes with $J \leq Q^{3/2}$ only in this paper.

\subsection{Images}

\begin{figure}[ht!]
\centering
\includegraphics[width=0.25\linewidth]{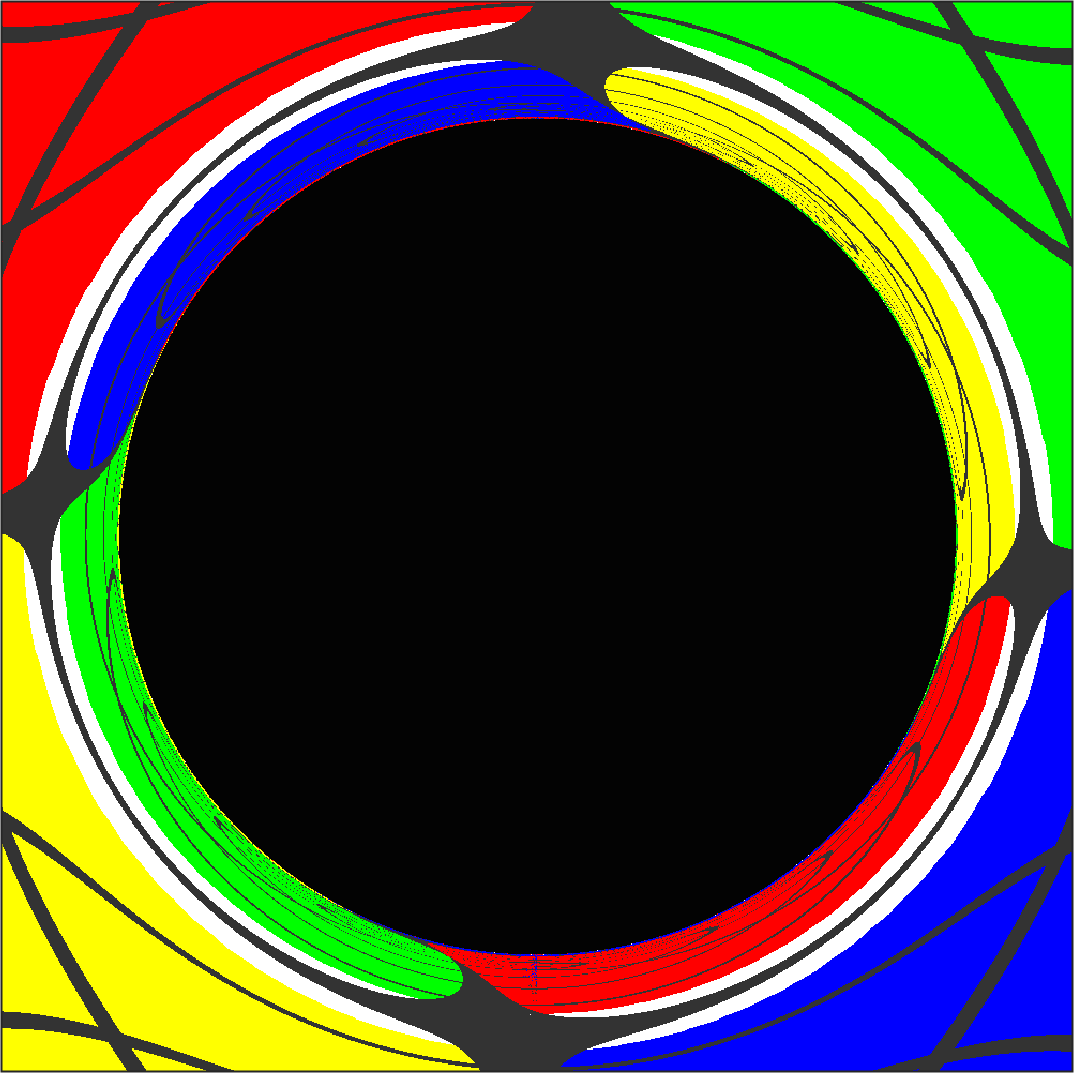} 
\includegraphics[width=0.25\linewidth]{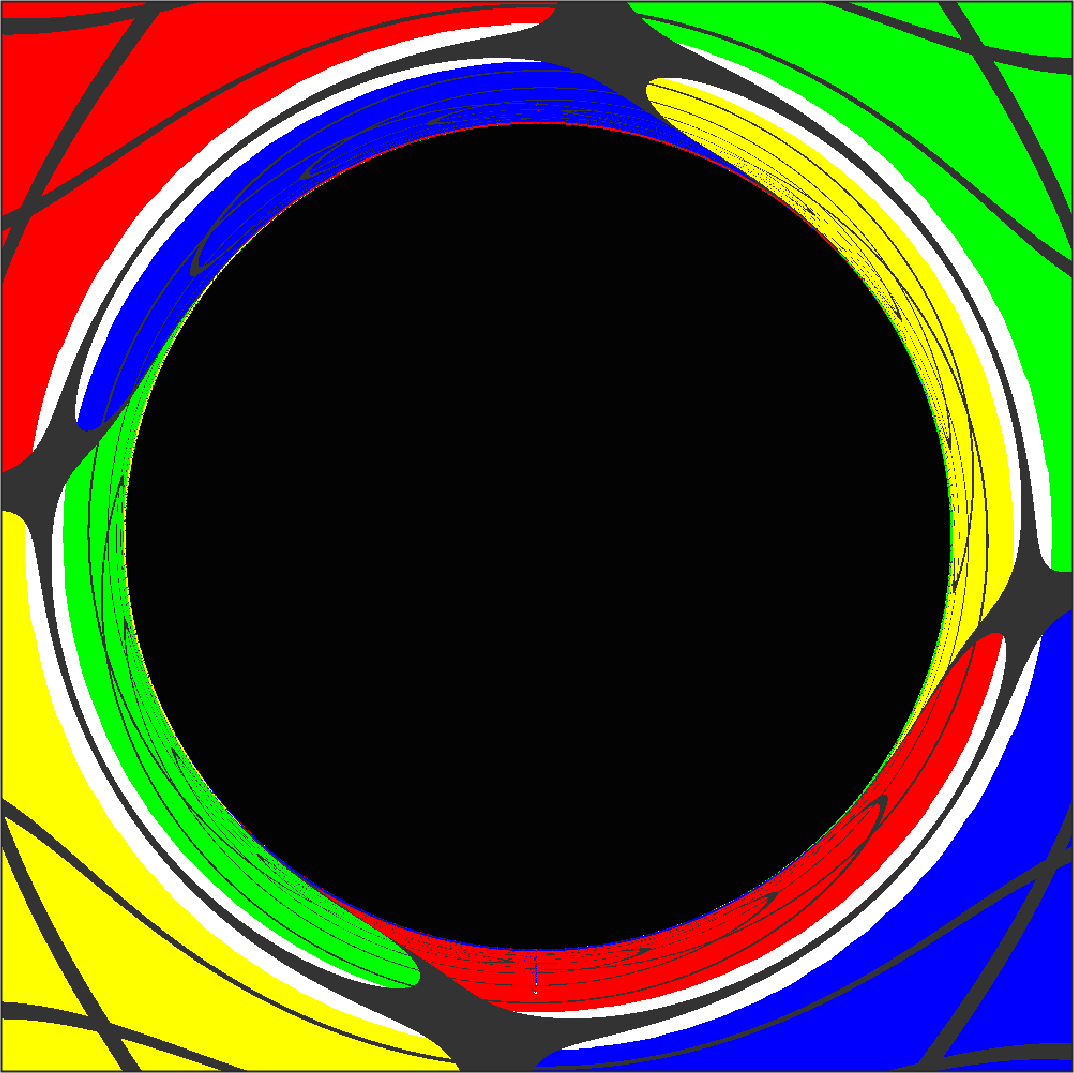} 
\includegraphics[width=0.25\linewidth]{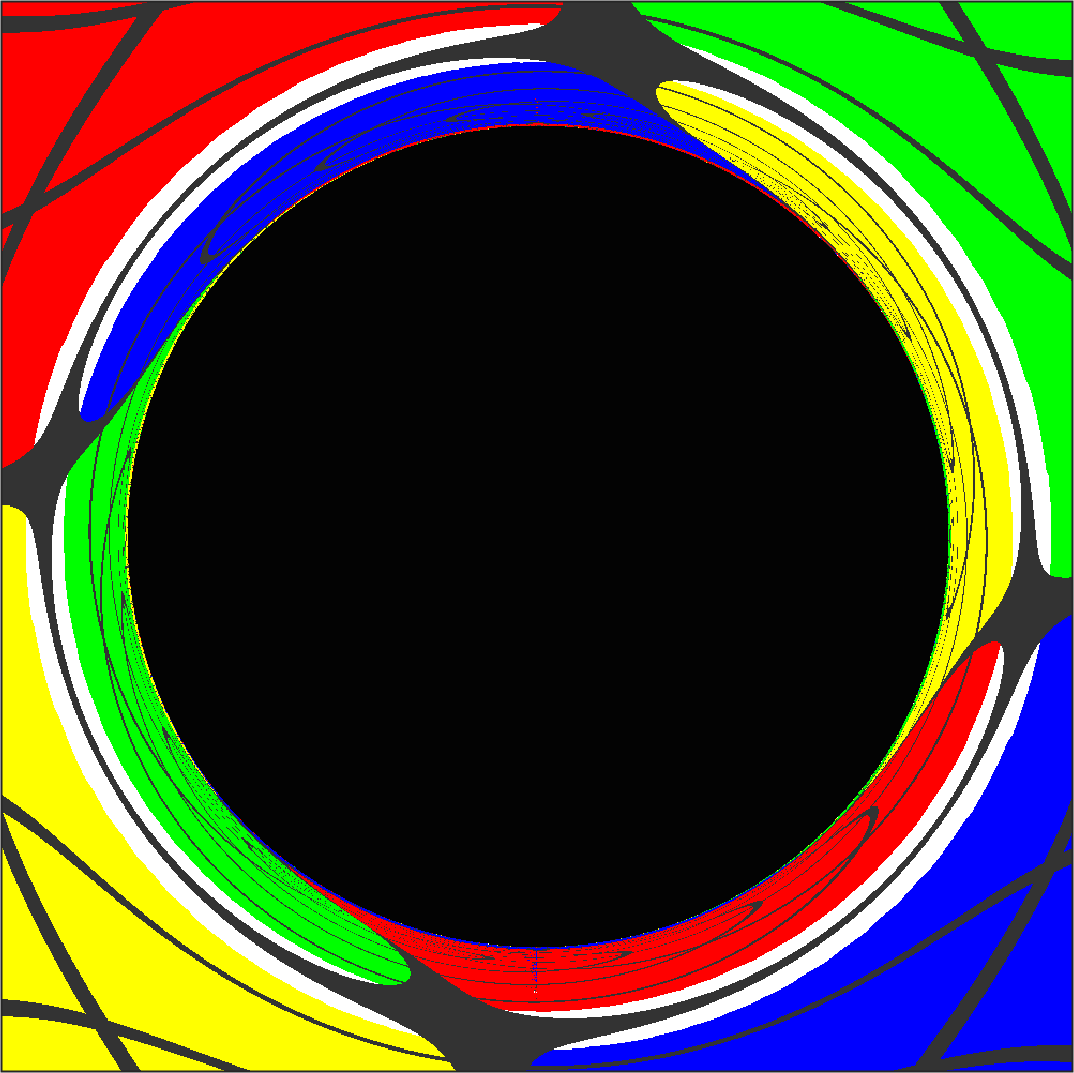} \\
\includegraphics[width=0.25\linewidth]{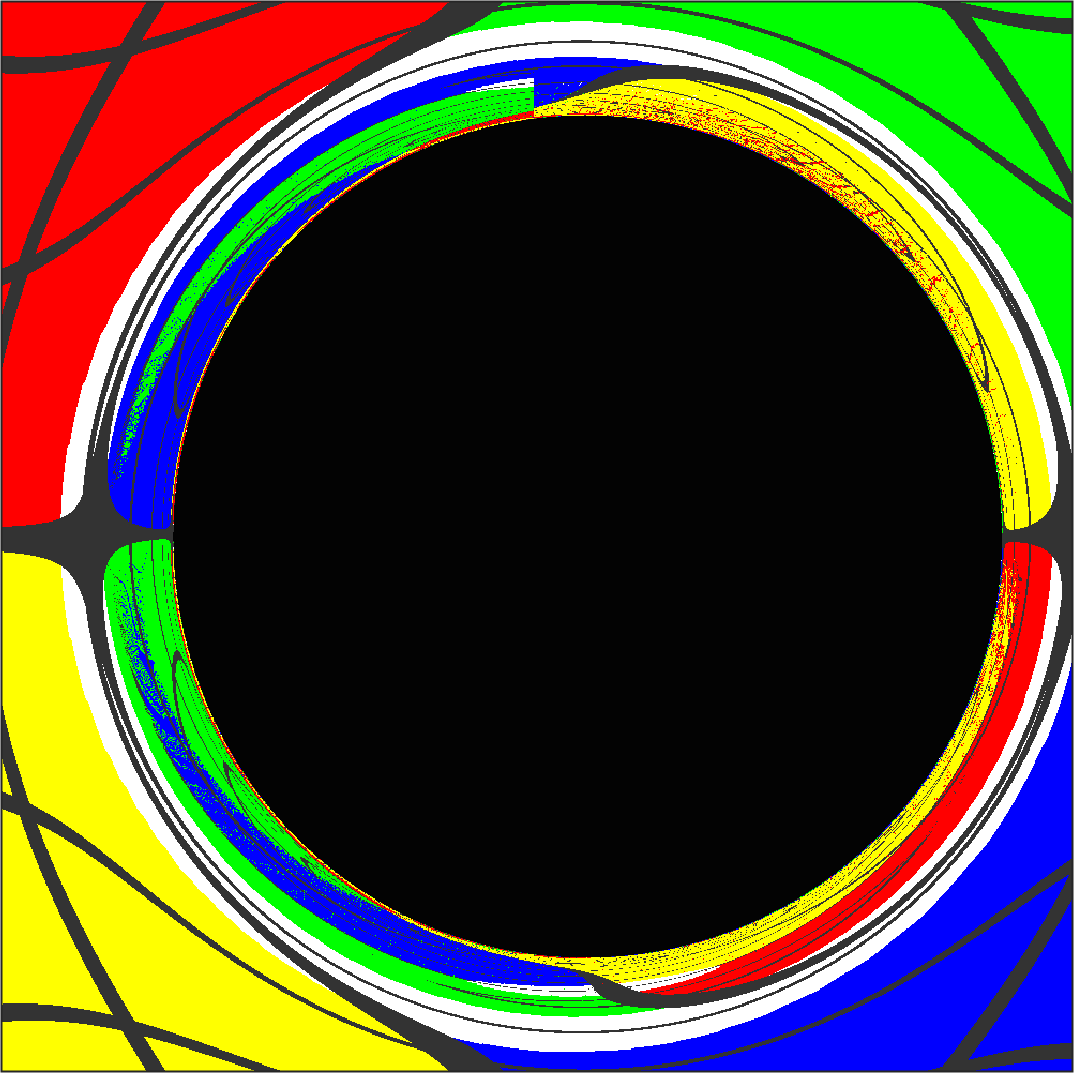} 
\includegraphics[width=0.25\linewidth]{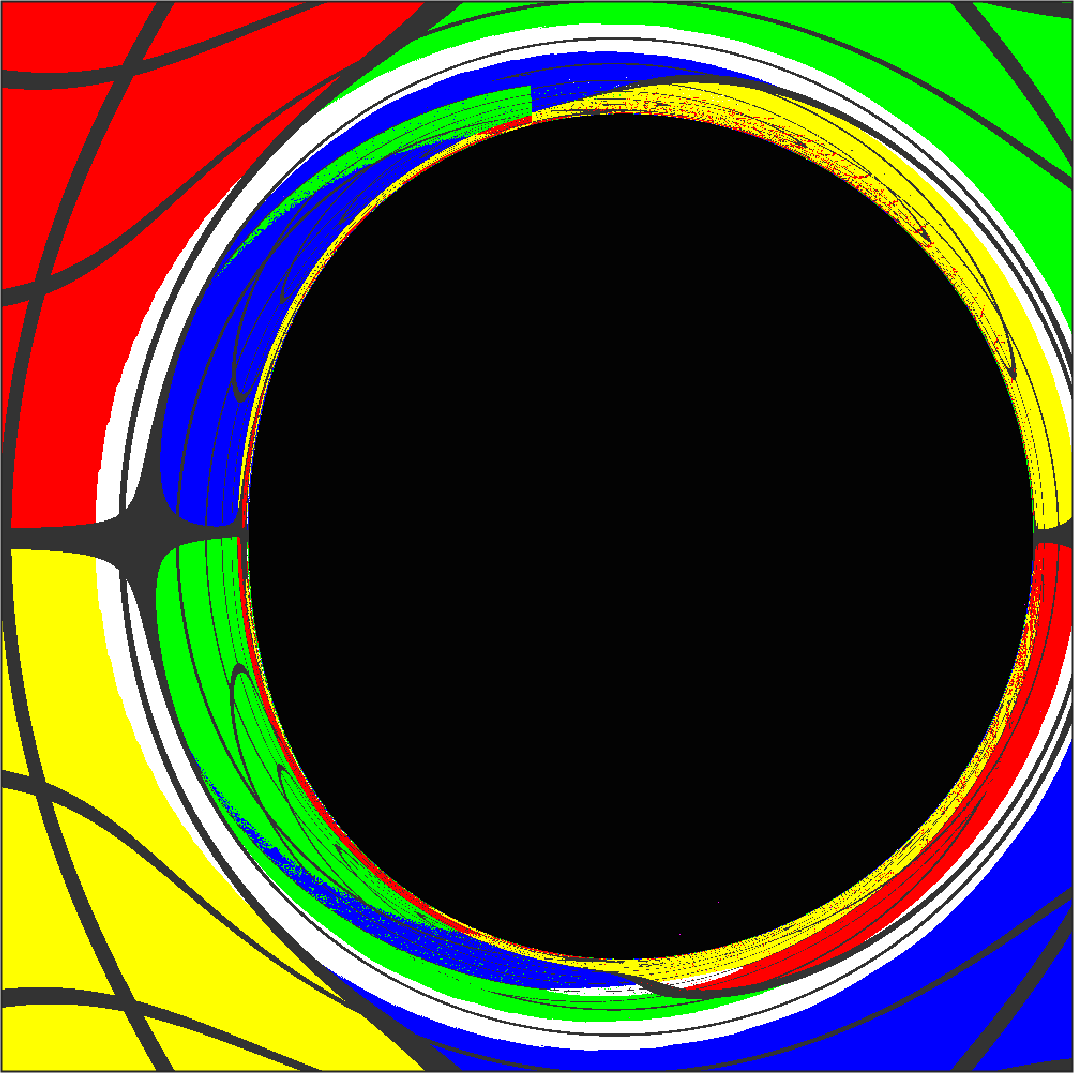} 
\includegraphics[width=0.25\linewidth]{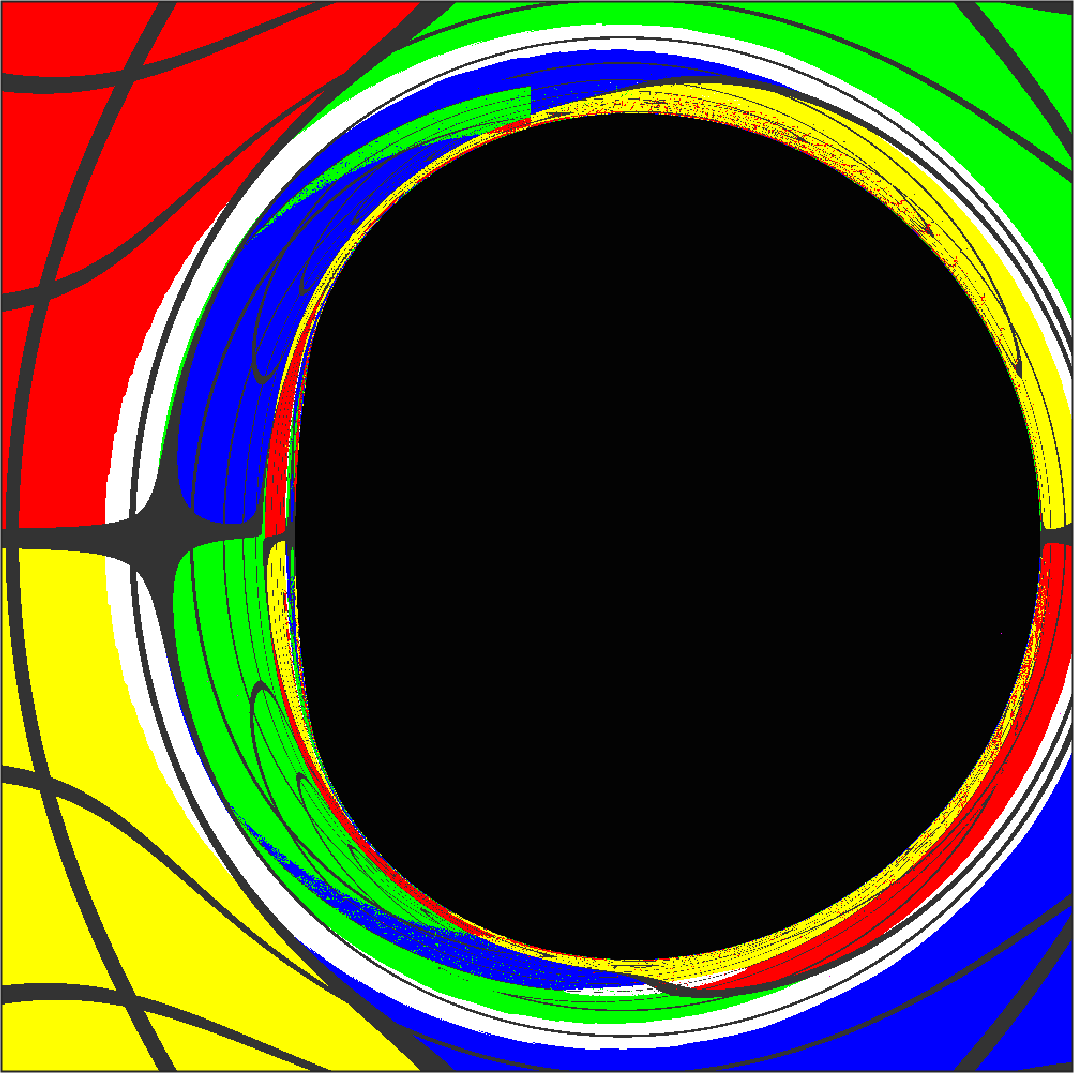} \\	
\includegraphics[width=0.25\linewidth]{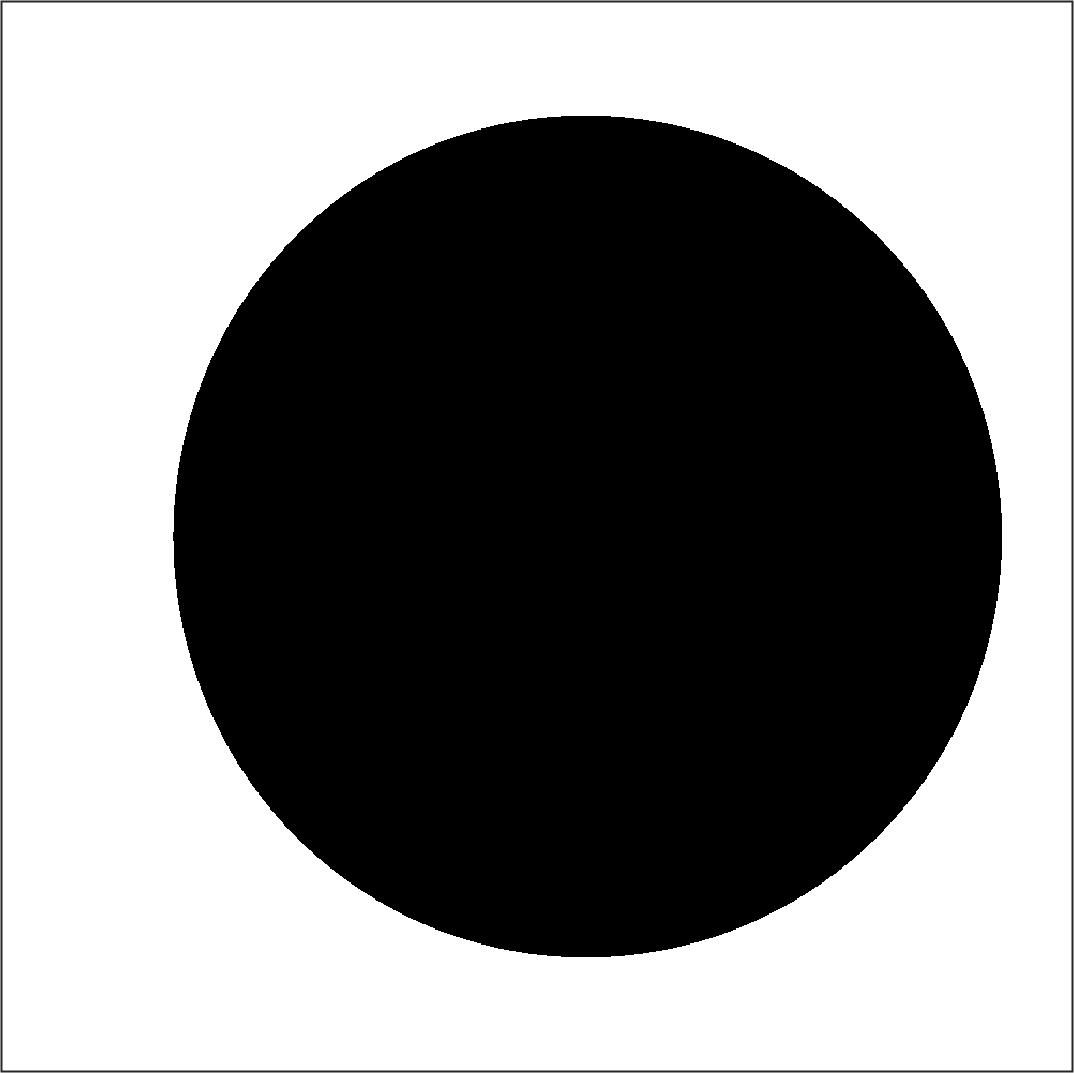} 
\includegraphics[width=0.25\linewidth]{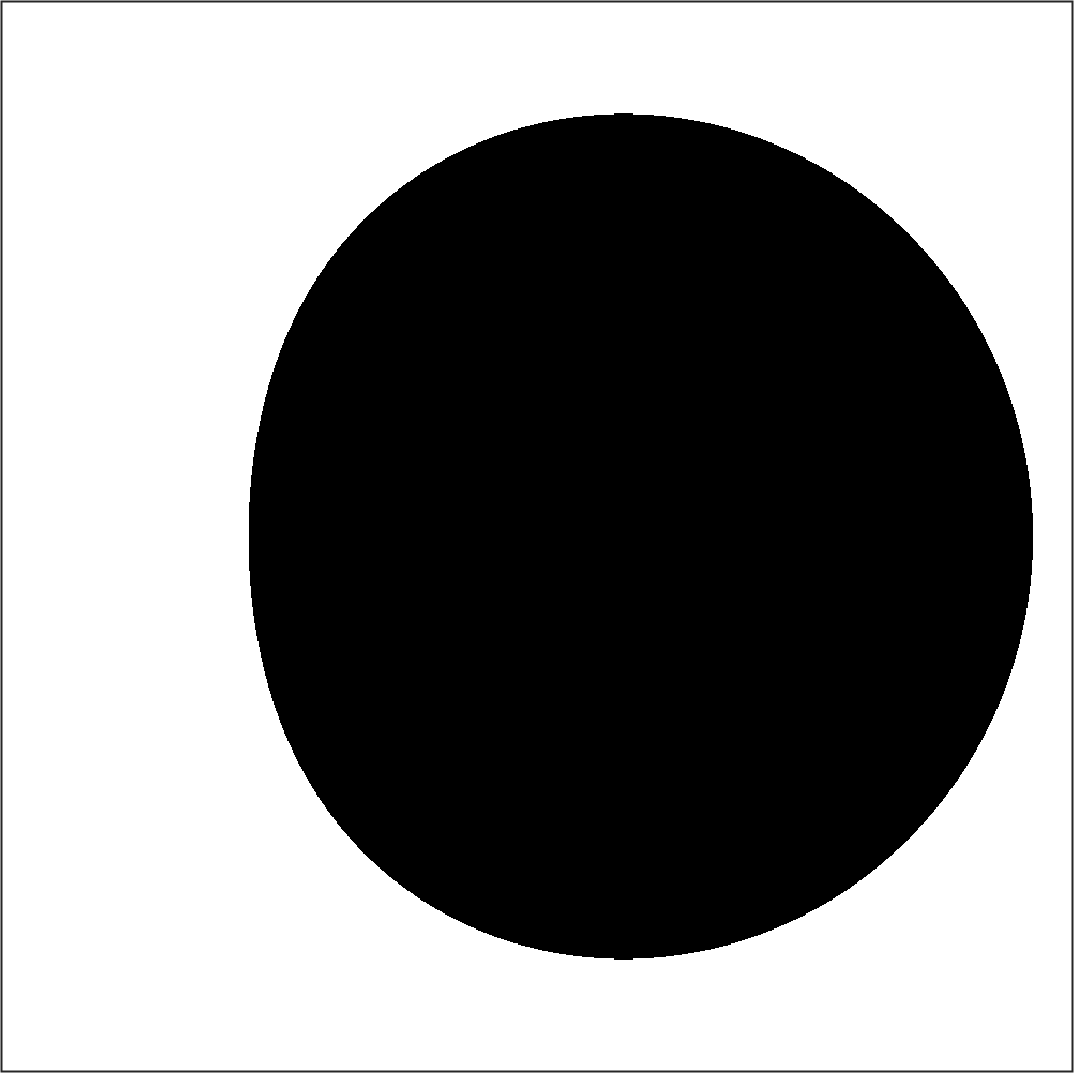} 
\includegraphics[width=0.25\linewidth]{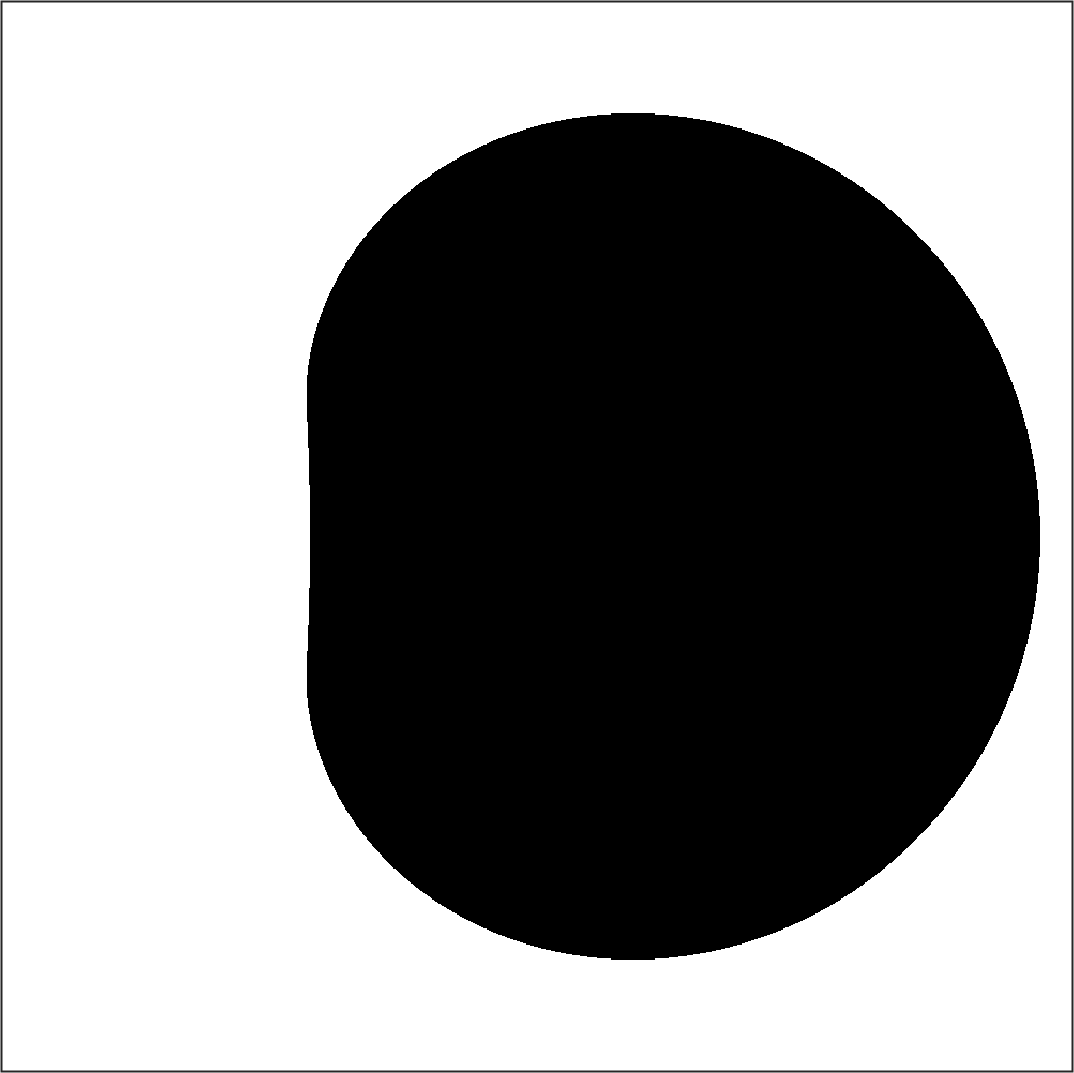} \\	
 \caption{Images of BMPV black holes for $x^4 =0$. From left to right: $J/Q^{3/2} = 0.5,0.9,1$. Top row: face-on, rotation perpendicular to line of sight.  Middle row: edge-on, rotation along line of sight. Bottom row: analytical shadow corresponding to second row.}
 \label{fig:BMPV_numeric}
\end{figure}

For imaging purposes, we will always suppress one dimension by fixing one of the four Cartesian spatial coordinates $x^1,x^2,x^3,x^4$ to zero. 
To clarify the basic physics picture, we choose inclination angle $i = 0$ and suppress one direction. By symmetries of the black hole metric, there are two interesting choices for the remaining 3 axes. Either there is non-zero angular momentum in the horizontal plane containing line of sight (``edge on''), or we see non-zero angular momentum in the plane perpendicular to the line of sight (``face on''). We present the images for 3 different values of the rotation parameter $J/Q^{3/2}$ in figure \ref{fig:BMPV_numeric}.

The images of the shadow show very similar behaviour as known from the rotating Kerr solution. Even though the horizon itself of the BMPV black hole is a  non-rotating null hypersurface \cite{Gauntlett:1998fz}, we do not see this effect in the shadow, since it images the light ring rather than the event horizon. As the angular momentum is increased, the shadow center shifts to the right, while the left side develops a flat side---the characteristic D-shape image known for rotating Kerr. This is due to the frame-dragging (on the left side of the image, rotation is oriented out of page). The edge-on view, with rotation perpendicular to line of sight, only shows the frame dragging in the image outside the round shadow.

We end this section with a comparison to analytic results. Since the geodesic problem in the BMPV background is Liouville integrable \cite{Gibbons:1999uv}, one can separate the geodesic and wave equations \cite{Herdeiro:2000ap,Diemer:2013fza}. We can then can obtain the equation for the edge of the black hole shadow following the same method as for the Kerr black hole. In appendix \ref{sec:BMPV_exact} we perform those calculations for the three-charge generalization of the BMPV black hole and discuss the equation for the shadow edge in terms of the impact parameters and inclination angle.

We found agreement between numerical and analytic results for the shadow edge  up to numerical precision for our chosen resolution. 
For angular momentum close to the extremal value  $0.99\lesssim J/Q^{3/2} < 1$ the numerical integration near the shadow edge is more error-prone due to the inner and outer horizon nearly coinciding. We still find agreement within our image resolution. For reference, we plot the analytic shadow in the bottom row of figure \ref{fig:BMPV_numeric}.

\section{Supersymmetric Black Rings}\label{sec:susyBR}

Our second example is the supersymmetric black ring \cite{Elvang:2004rt,Bena:2004de}. 

\subsection{Metric}

We write the metric again in the form
\begin{equation}
 ds^2 = - Z^{-2/3} (dt + k)^2 + Z^{-1/3} ds^2_4 (\mathbb{R}^4),
\end{equation}
with the coordinates \eqref{eq:GH_R4}. The black ring has \emph{two} centers (two poles in harmonic functions): one located at the origin of $\mathbb{R}^3$, the other we put at the positive $z$-axis at a distance $a$. This is reflected in the form of $Z$ and $k$:
\begin{IEEEeqnarray}{rCl}
 Z &=& 1 + \frac{Q-q^2}{4 \Sigma} + \frac{q^2 r}{4 \Sigma^2},\\
k &=& \left[\left(\frac{q^3 r^2 }{8 \Sigma^3}-\frac{3 q r  \left(q^2-Q\right)}{16 \Sigma^2}\right)(1+\cos (\theta ))+\frac{3 q \left(4 r+R^2\right)}{8  \Sigma}-\frac{3 q}{2}\right] \dif \phi \nonumber\\
&& + \left[\frac{q^3 r^2}{8 \Sigma^3}-\frac{3 q r \left(q^2-Q\right)}{16 \Sigma^2}+\frac{3 q \left(4 r+R^2\right)}{16 \Sigma}-\frac{3 q}{4}\right] \dif \psi,
\end{IEEEeqnarray}
with $\Sigma = \sqrt{r^2 - 2 a r \cos \theta + a^2}$ the coordinate distance of the ring location to the origin.

\subsection{Physical Properties}

Again this solution has an electric charge $Q$, but unlike the black hole it has two independent angular momenta. The origin of this freedom is the presence of a dipole charge $q$, related to non-trivial magnetic fields:
\begin{equation}
M = \frac{4 G_{5}}{\pi} Q\,,\qquad J_{12} - J_{34} = \frac {3\pi} {G_5}a q\,.
\end{equation}
Again we will put $\frac \pi {4 G_5} =1$.
The black ring has non-spherical horizon topology: the horizon has the topology of $S^1 \times S^2$, a higher-dimensional torus with $S^2$ cross-section. The horizon area of the ring is the product of the $S^1$ length, with radius $L = \frac{\sqrt{3}}{2q}\sqrt{(Q-q^2)^2-16a q^2}$, and the $S^2$ area, which has radius $q/2$:
\begin{equation}
 A_H = 2 \pi^2 q^2 L >0 \,.
\end{equation}

\begin{figure}[ht!]
\centering
\includegraphics[width=.75\textwidth]{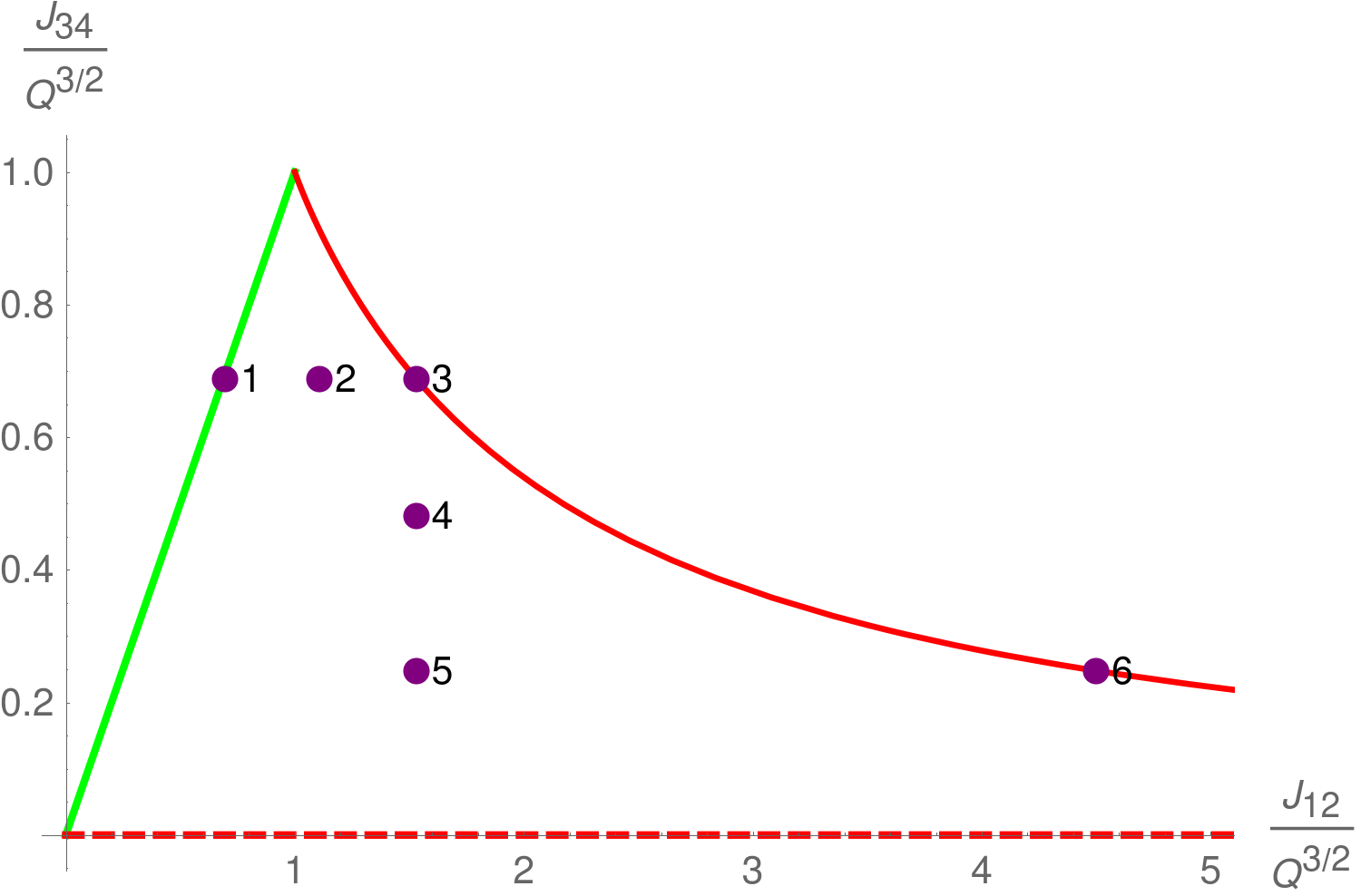}
\caption{The phase diagram of black holes and black rings in the $(J_{12},J_{34})$-plane. Angular momenta are normalized  by $Q^{3/2}$.  Black holes correspond to the green diagonal line, black rings lie under the curved red line and extend along the horizontal axis to unbounded $J_{12}$. The purple dots show particular solutions we image below. 
}
\label{fig:phasediagram1}
\end{figure}

As for the black hole, the requirement of positive horizon area puts a bound on the angular momenta. However, this is an insufficient condition for physical regularity. Absence of CTC's requires also that $0 < q <\sqrt{Q}$. In addition the ring radius parameter $a$ should be positive. We plot the region of allowed ring solutions in the ($J_{12},J_{34})$ plane in figure \ref{fig:phasediagram1}. We choose to normalize the angular momenta by an appropriate power of the charge $Q$. This is not a mere fixing of scale, but is a judicious choice that makes use of the  invariance of the spectrum of solutions of 5D supergravity coupled to vector multiplets under the rescaling:
\begin{equation}
(Q, J_{12}, J_{34}) \to (\lambda^2 Q, \lambda^3 J_{12}, \lambda^3 J_{34}) \,,\quad \lambda>0\,,
\end{equation}
which map solutions to solutions. Hence the two-dimensional diagram truly represents all solutions in this three-parameter class of black rings.

Depending on the ratio of ring circumference by cross-sectional area, set by $L/q$, we can distinguish two distinct types of black rings. When that ratio is close to its minimal value $L/q \gtrsim 2$, we have a `thick ring', shaped like a higher-dimensional donut with a relatively small hole in the middle. For  $L/q \gg 2$ we have a `thin ring'. One can check that thick rings have values of angular momenta that are nearly equal (close to the black hole vlaues), while the thinnest rings have $J_{12} \gg J_{34}$. We will see below that indeed thin rings generate quite a different image than black holes.

As a final remark, we note that although classically the charges take on a continuous range of values, charge quantization in string theory requires $Q$ and $q$ to take on discrete values.

\subsection{Images}

\begin{figure}[ht!]
\centering
\begin{tabular}{c|c|c|c}

\resizebox {0.22\linewidth} {!} {
	\begin{tikzpicture}
	
	\begin{scope}[canvas is xz plane at y=0]
	\draw [->] (0,0) -- (0,1.5) node[left] {$x_{1}$};
	\draw [->] (0,0) -- (1.5,0) node[above] {$x_{4}$};
	\end{scope}	
	\draw [->] (0,0) -- (0,1.5) node[above] {$x_{3}$};
	
	\begin{scope}[canvas is xy plane at z=1]
	\filldraw (0,0) circle (0.15);
	\end{scope}
	\begin{scope}[canvas is xy plane at z=-1]
	\filldraw (0,0) circle (0.15);
	\end{scope}
	
	\begin{scope}[canvas is xy plane at z=0]
	\draw[thick] [->] (10:1.25) arc (10:80:1.25) node[midway,above,xshift=0.1cm,yshift=0.1cm] {$J_{34}$};
	\end{scope}
	
	\begin{scope}[canvas is yz plane at x=0,shift={(3.5,0)},rotate=-180,scale=0.75]
	\filldraw [black] (0,0) circle (2pt);
	\draw (0.9396926208,-0.3420201433) arc (-20:20:1);
	\draw (1.409538931,-0.513030215) -- ++(-20+180:1.5) -- +(20:1.5);
	\node at (0,0)[left]{O};
	\end{scope}
	
	\end{tikzpicture}} &

\resizebox {0.22\linewidth} {!} {
	\begin{tikzpicture}
	
	\begin{scope}[canvas is xz plane at y=0]
	\draw [->] (0,0) -- (0,1.5) node[left] {$x_{1}$};
	\draw [->] (0,0) -- (1.5,0) node[above] {$x_{4}$};
	\end{scope}	
	\draw [->] (0,0) -- (0,1.5) node[above] {$x_{3}$};
	
	\begin{scope}[canvas is xy plane at z=1]
	\filldraw (0,0) circle (0.15);
	\end{scope}
	\begin{scope}[canvas is xy plane at z=-1]
	\filldraw (0,0) circle (0.15);
	\end{scope}
	
	\begin{scope}[canvas is xy plane at z=0]
	\draw[thick] [->] (10:1.25) arc (10:80:1.25)node[midway,above,xshift=0.1cm,yshift=0.1cm] {$J_{34}$};
	\end{scope}
	
	\begin{scope}[canvas is yz plane at x=0,shift={(0,4)},rotate=-90,scale=0.75]
	\filldraw [black] (0,0) circle (2pt);
	\draw (0.9396926208,-0.3420201433) arc (-20:20:1);
	\draw (1.409538931,-0.513030215) -- ++(-20+180:1.5) -- +(20:1.5);
	\node at (0,0)[left]{O};
	\end{scope}
	
	\end{tikzpicture}} &

\resizebox {0.22\linewidth} {!} {	
	\begin{tikzpicture}
	
	\begin{scope}[canvas is xz plane at y=0]
	\draw [->] (0,0) -- (0,1.5) node[left] {$x_{1}$};
	\draw [->] (0,0) -- (1.5,0) node[above] {$x_{2}$};
	\end{scope}	
	\draw [->] (0,0) -- (0,1.5) node[above] {$x_{3}$};
	
	\begin{scope}[canvas is xz plane at y=0]
	\draw[ultra thick] (0,0) circle (1cm);
	\draw[thick] [->] (10:1.4) arc (10:80:1.4)node[midway,below] {$J_{12}$};
	\end{scope}
	
	\begin{scope}[canvas is yz plane at x=0,shift={(3.5,0)},rotate=-180,scale=0.75]
	\filldraw [black] (0,0) circle (2pt);
	\draw (0.9396926208,-0.3420201433) arc (-20:20:1);
	\draw (1.409538931,-0.513030215) -- ++(-20+180:1.5) -- +(20:1.5);
	\node at (0,0)[left]{O};
	\end{scope}
	
	\end{tikzpicture}} &

\resizebox {0.22\linewidth} {!} {
	\begin{tikzpicture}
	
	\begin{scope}[canvas is xz plane at y=0]
	\draw [->] (0,0) -- (0,1.5) node[left] {$x_{1}$};
	\draw [->] (0,0) -- (1.5,0) node[above] {$x_{2}$};
	\end{scope}	
	\draw [->] (0,0) -- (0,1.5) node[above] {$x_{3}$};
	
	\begin{scope}[canvas is xz plane at y=0]
	\draw[ultra thick] (0,0) circle (1cm);
	\draw[thick] [->] (10:1.4) arc (10:80:1.4)node[midway,below] {$J_{12}$};
	\end{scope}
	
	\begin{scope}[canvas is yz plane at x=0,shift={(0,4)},rotate=-90,scale=0.75]
	\filldraw [black] (0,0) circle (2pt);
	\draw (0.9396926208,-0.3420201433) arc (-20:20:1);
	\draw (1.409538931,-0.513030215) -- ++(-20+180:1.5) -- +(20:1.5);
	\node at (0,0)[left]{O};
	\end{scope}
	
	\end{tikzpicture}} \\

\rule{0pt}{17ex}
\includegraphics[width=0.22\linewidth]{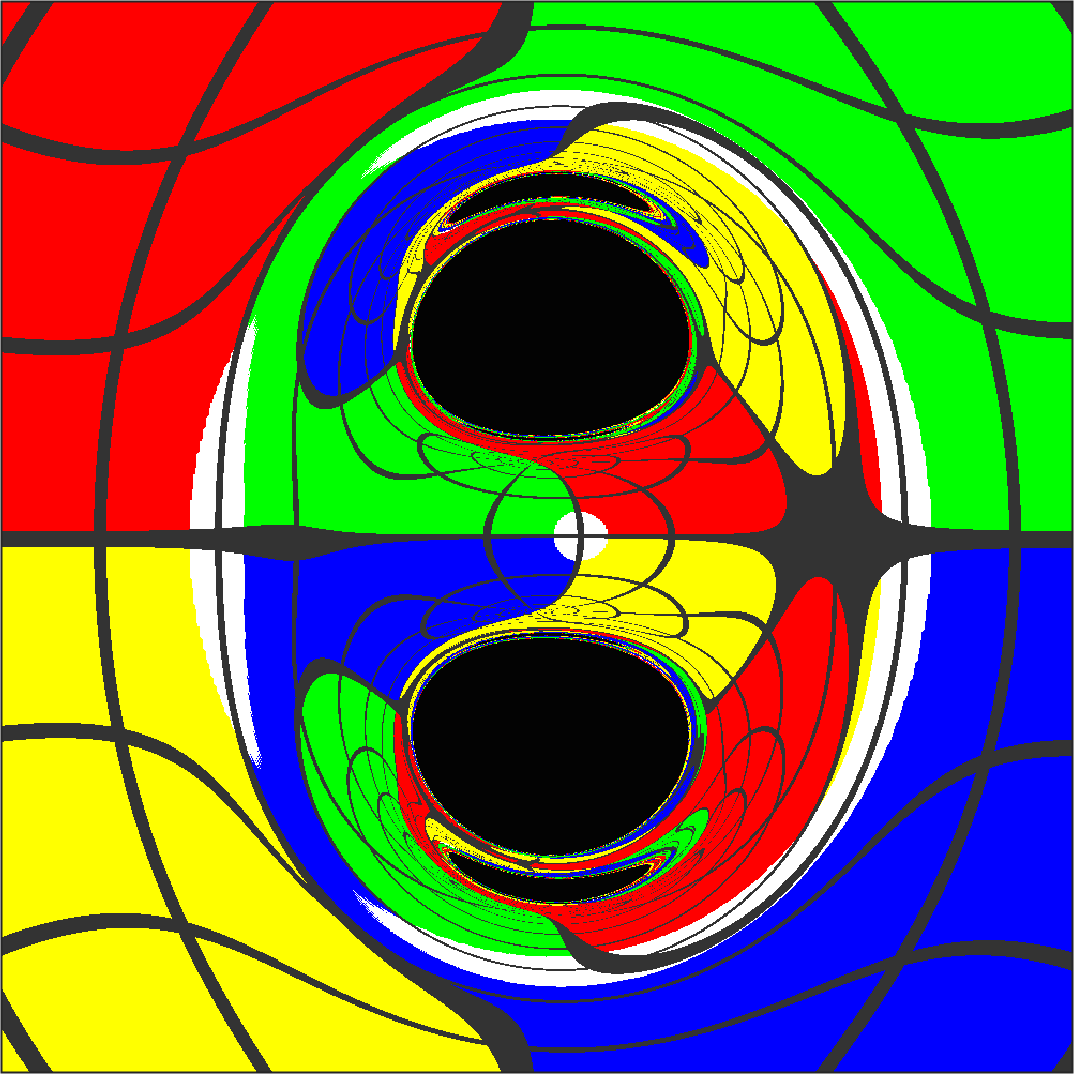} &
\includegraphics[width=0.22\linewidth]{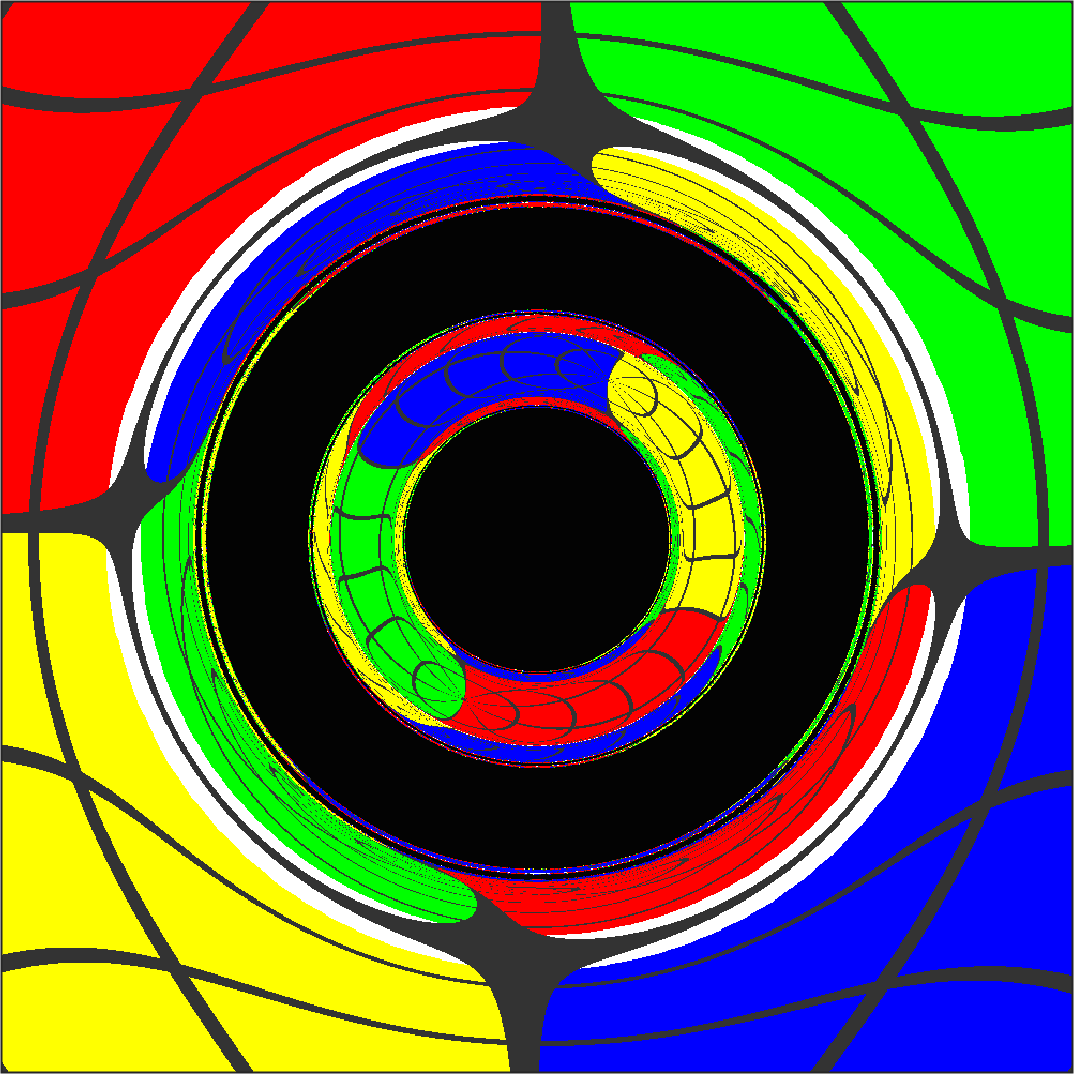} &
\includegraphics[width=0.22\linewidth]{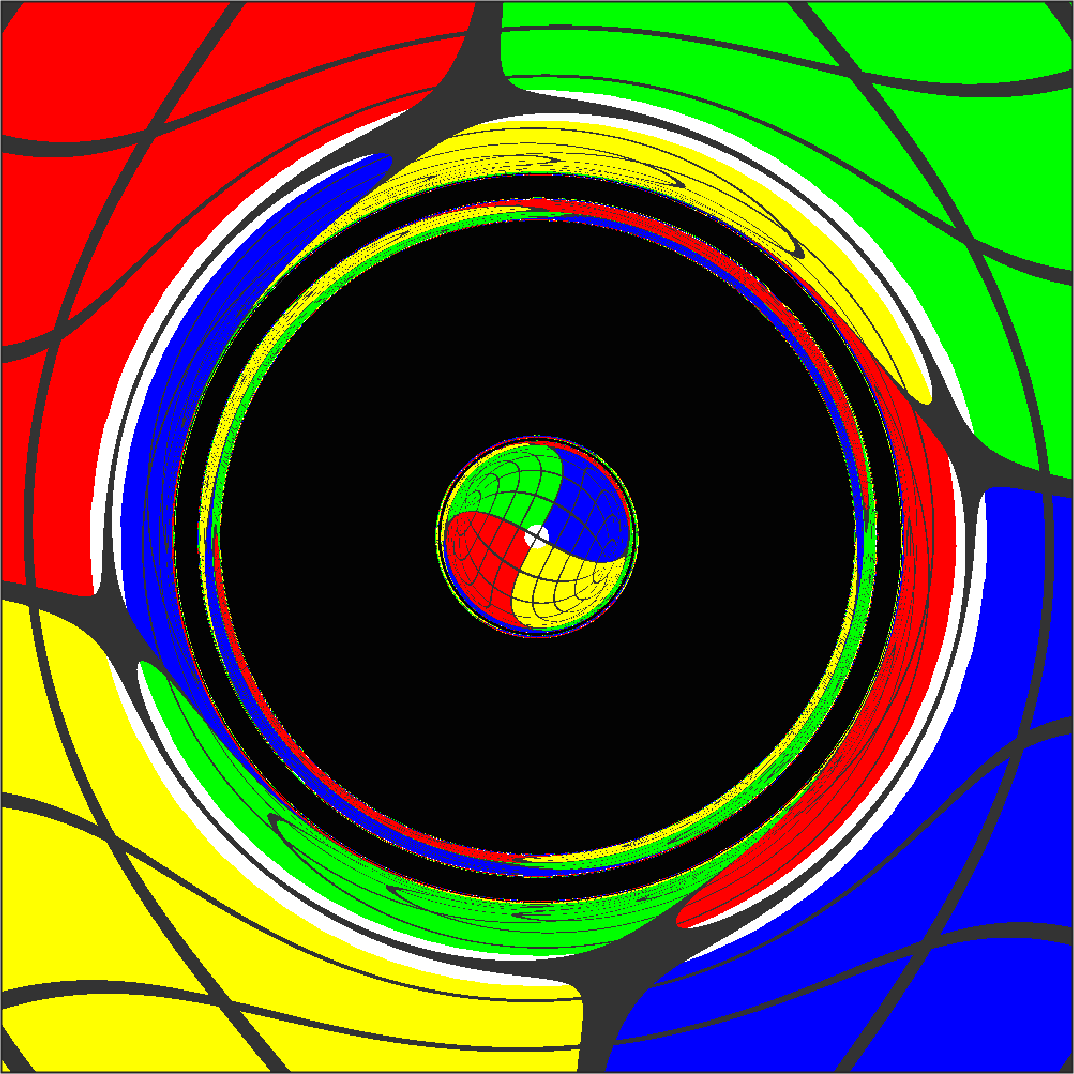} &
\includegraphics[width=0.22\linewidth]{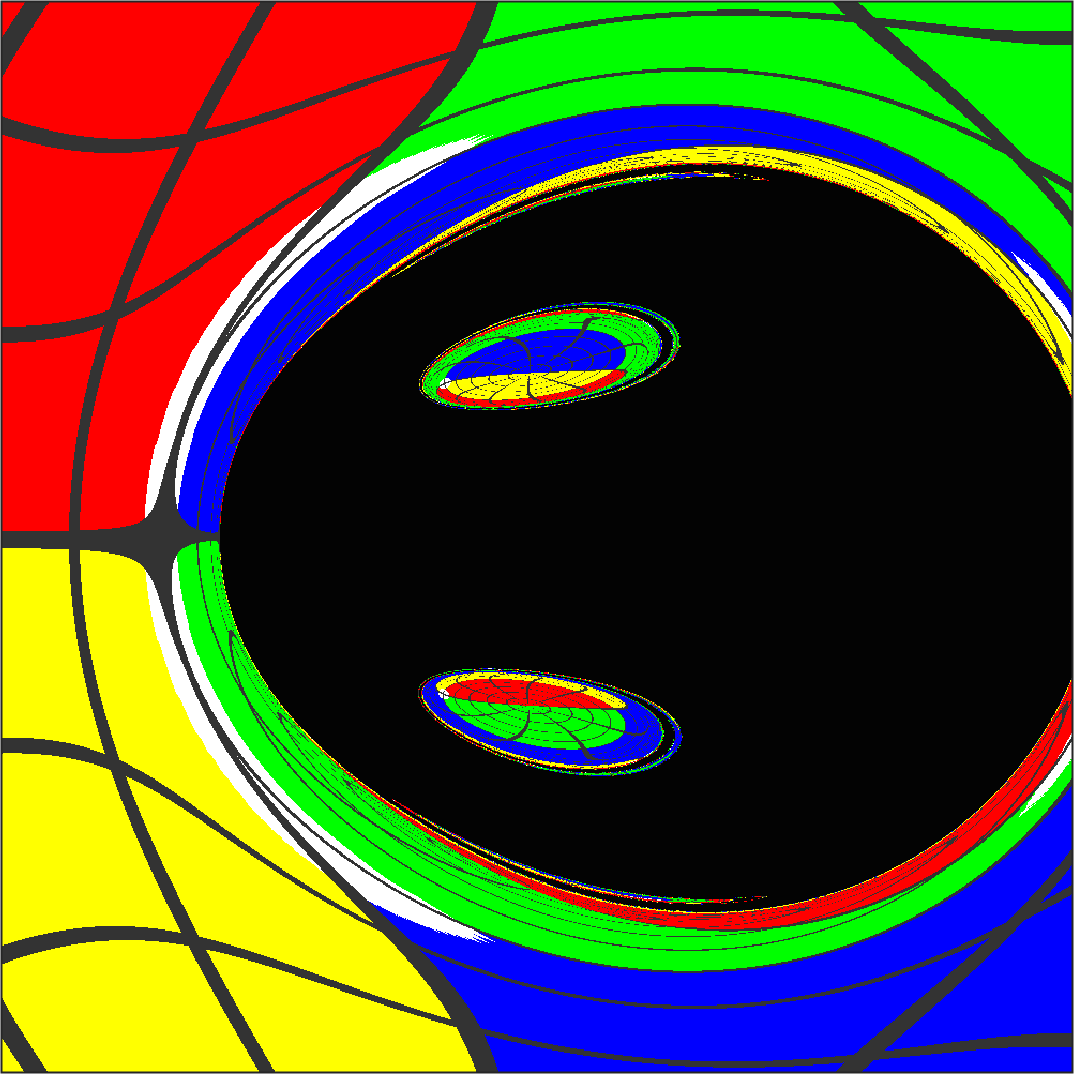}\\

\rule{0pt}{17ex}
\includegraphics[width=0.22\linewidth]{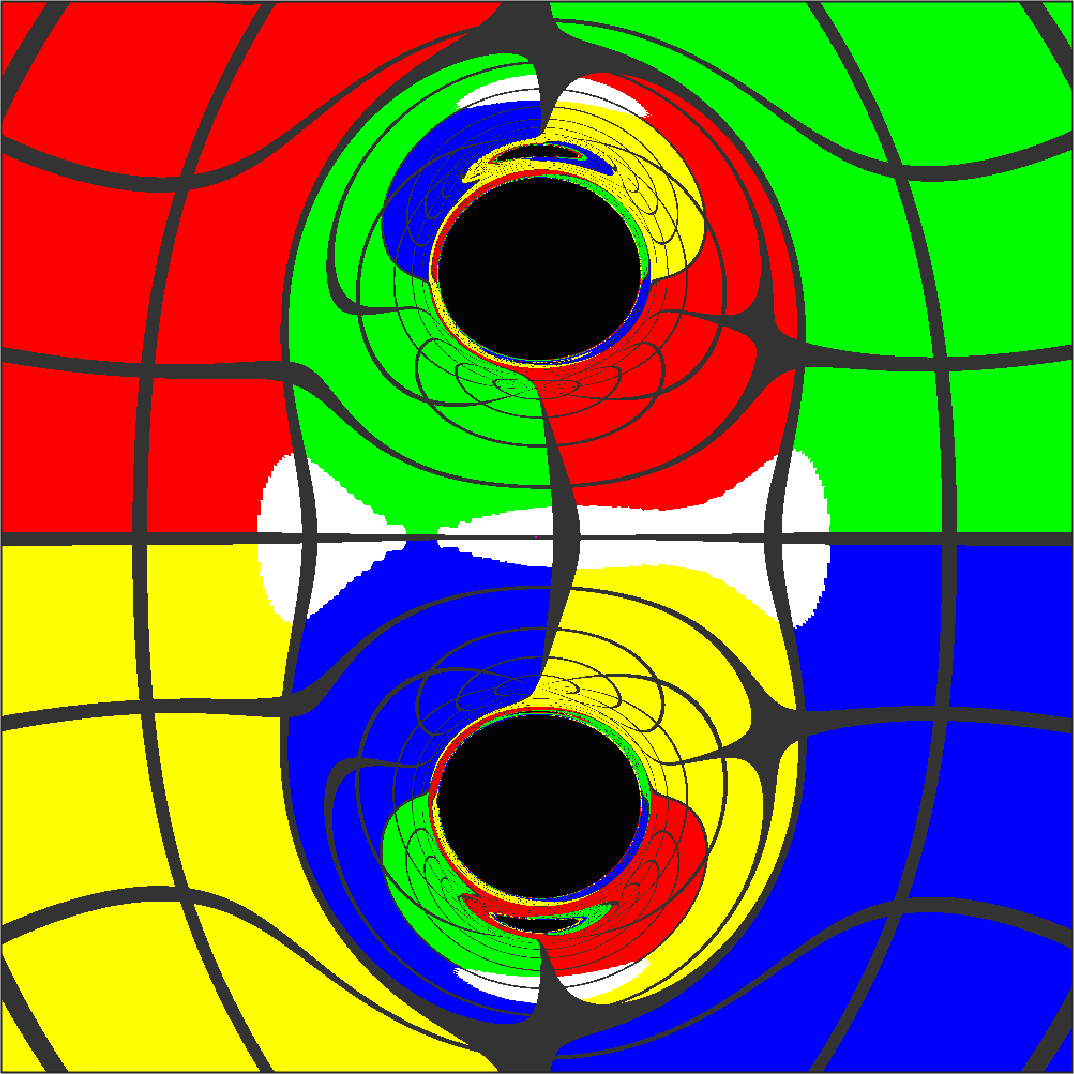} &
\includegraphics[width=0.22\linewidth]{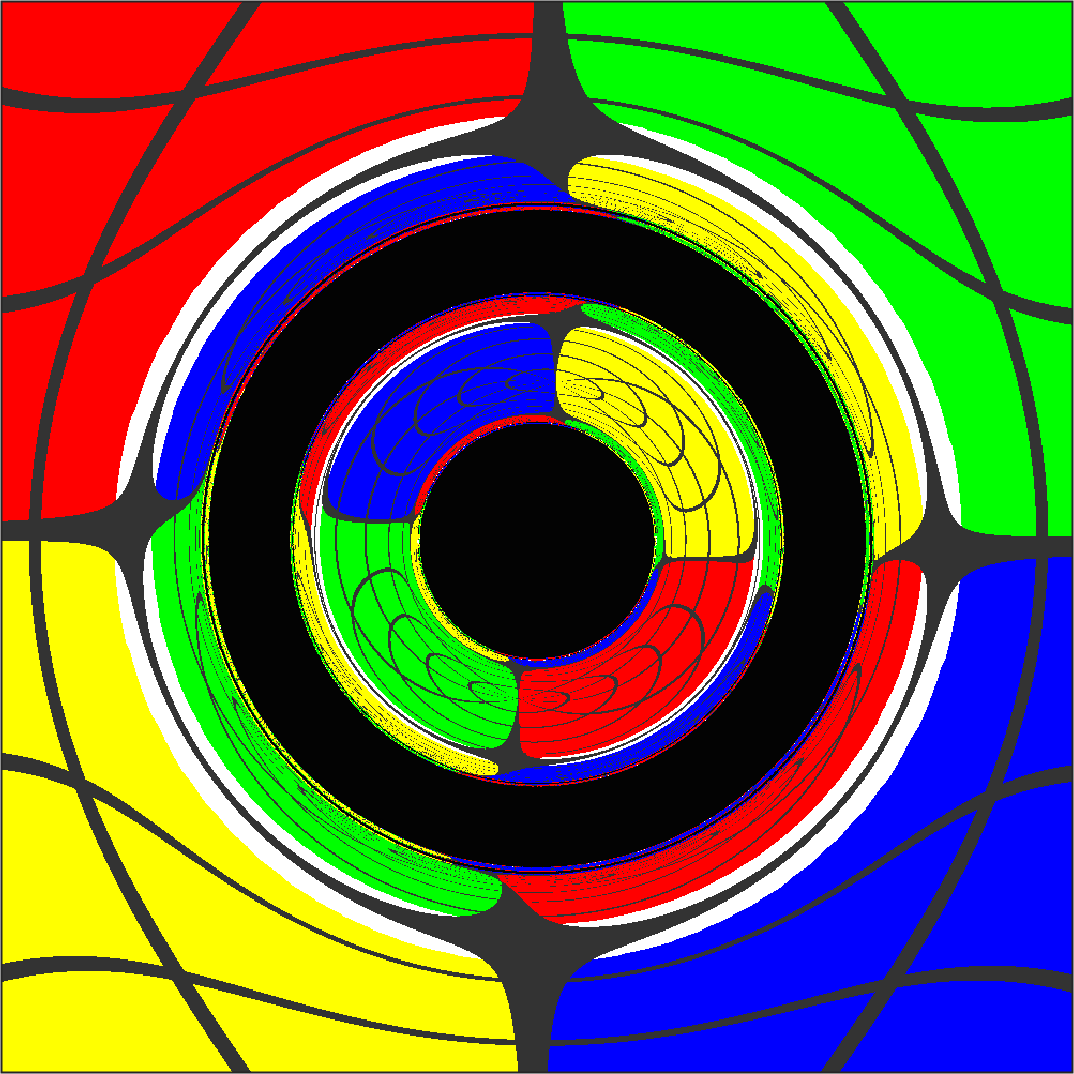} &
\includegraphics[width=0.22\linewidth]{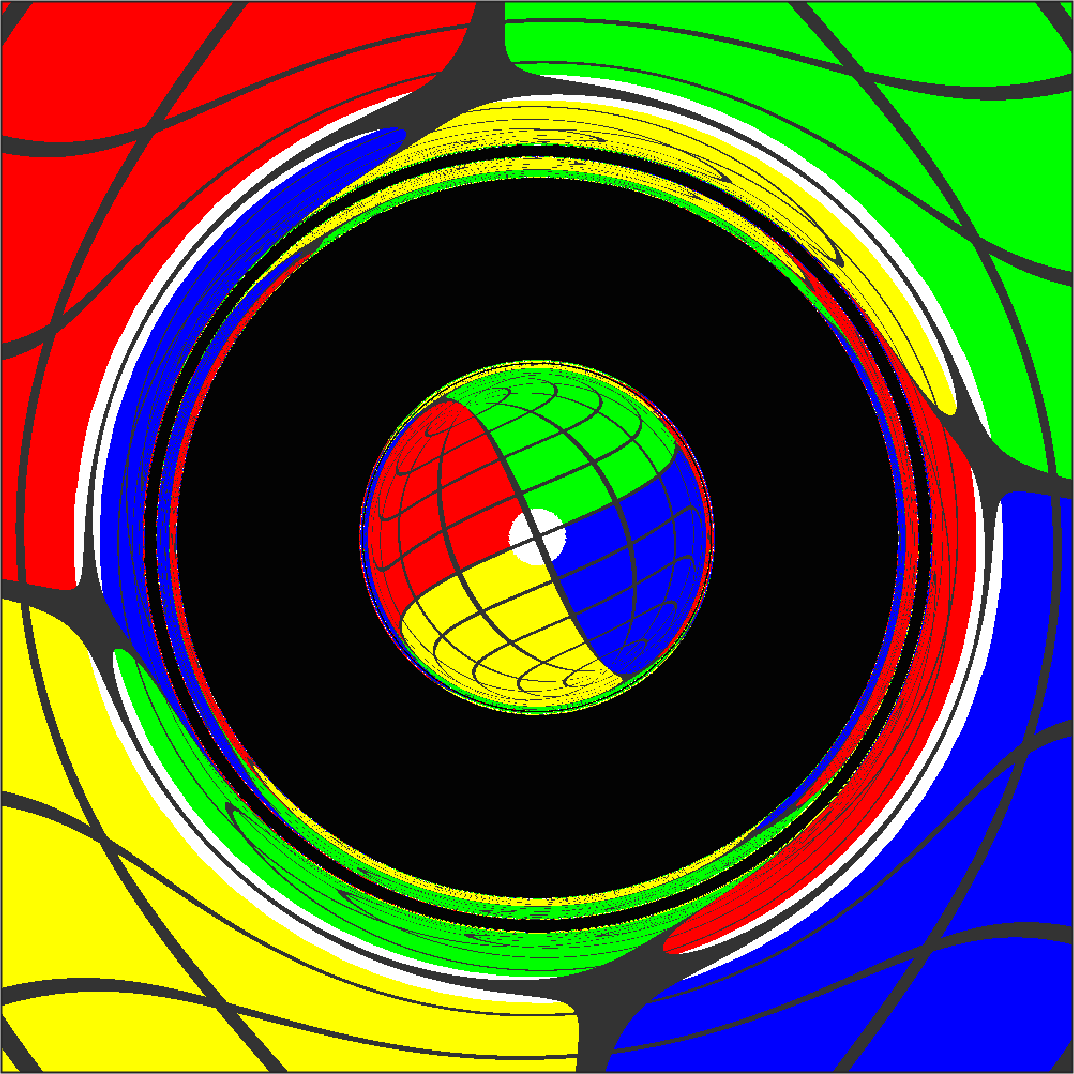} &
\includegraphics[width=0.22\linewidth]{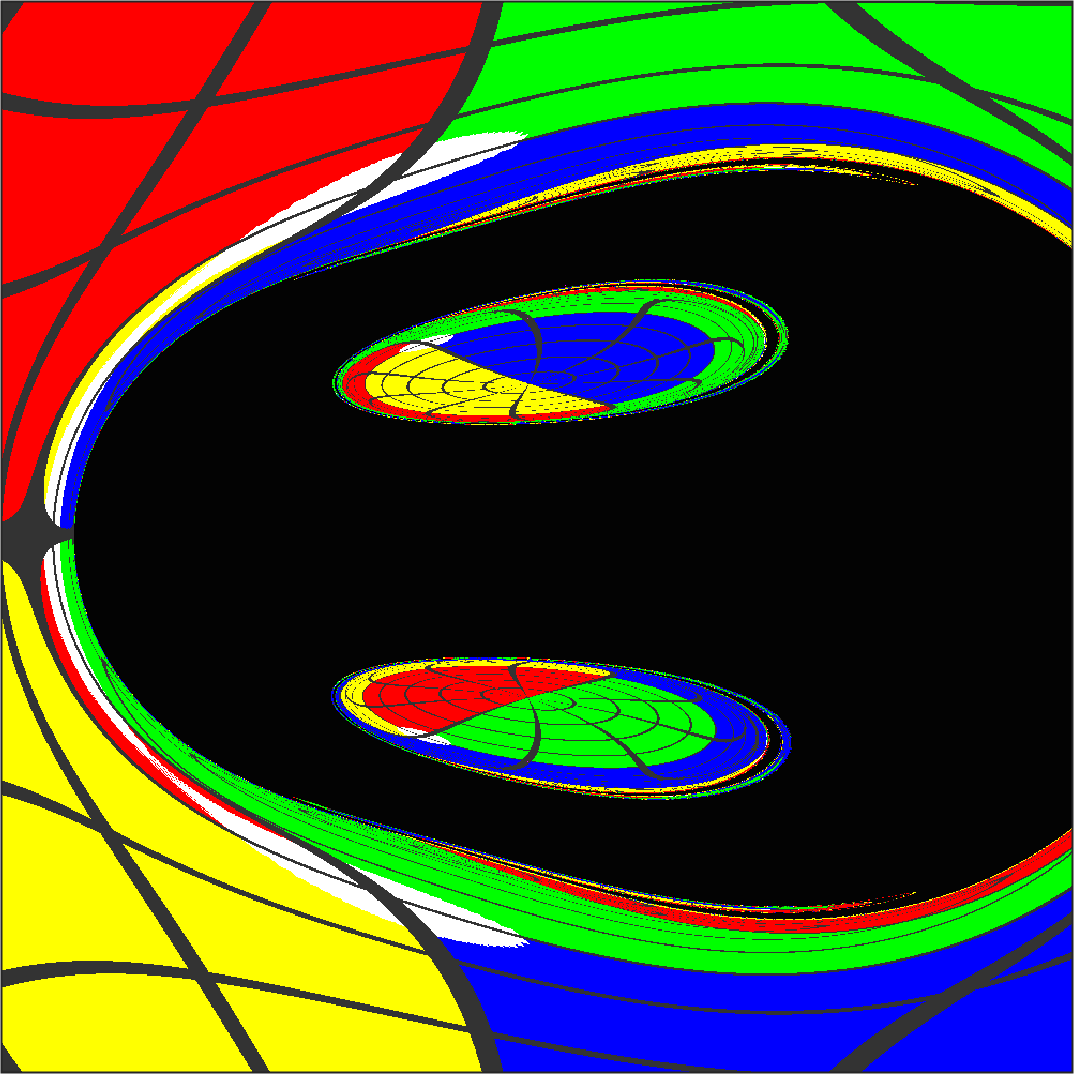}

\end{tabular}
\caption{Images of black rings with constant $J_{12} = \frac{49}{32} Q^{3/2}$ organized as in table \ref{tab:four_viewings}. Top row: the position of the camera for the four main types of images we provide in this paper.  Middle row: ring 4 in phase diagram of figure \ref{fig:phasediagram1}, with $J_{34}=\frac{13}{27}Q^{3/2}$. Bottom row: ring 5 with $ J_{34} = \frac{107}{432}Q^{3/2}$.}
\label{fig:four_viewings}
\end{figure}
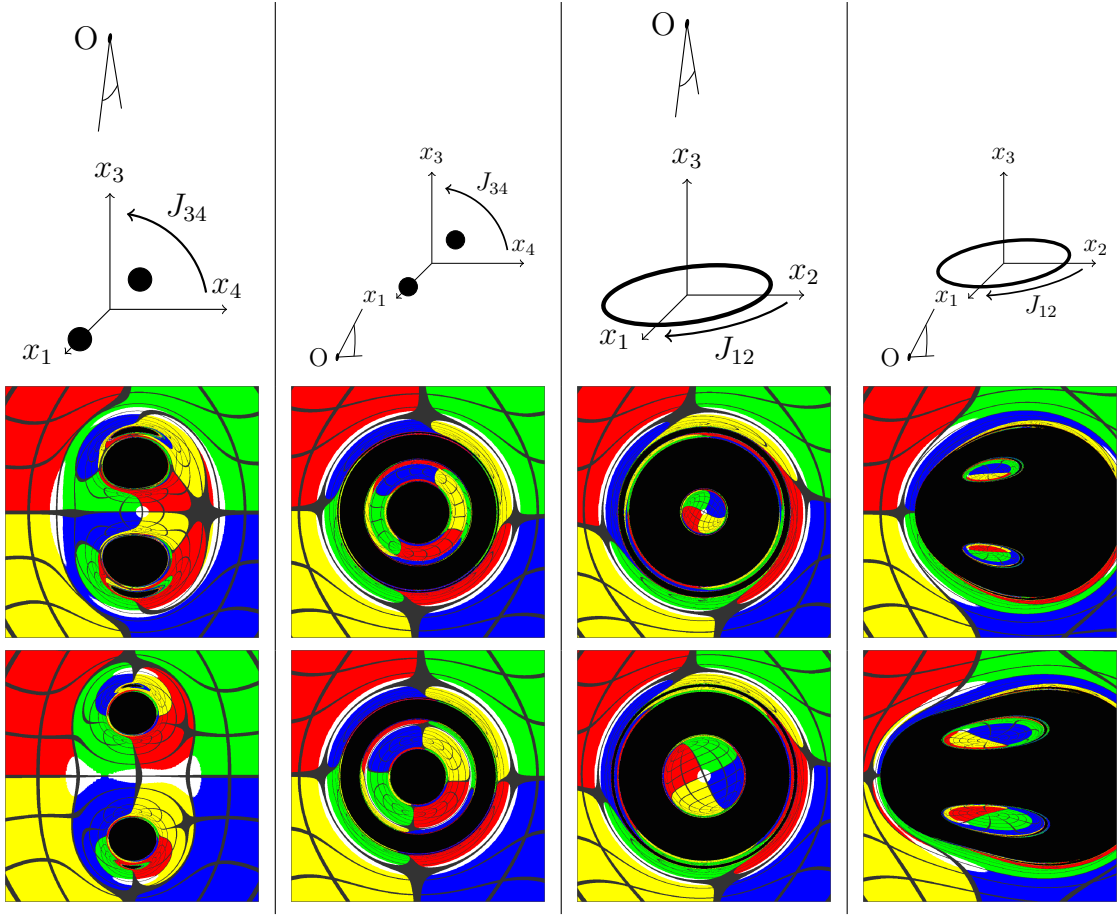

Unlike for the black hole, different slicings are now giving physically distinct two-dimensional images for a given viewing angle. Upon fixing one Cartesian direction, the two-dimensional projection of the $S^1 \times S^2$ horizon area will be either two disconnected $S^2$'s, or a two-dimensional torus with $S^1 \times S^1$ topology. For each projections we show two different viewing angles that we dub ``face-on'' and ``edge-on'', see table \ref{tab:four_viewings} for an overview. We will make most plots in the order of the table, see figure \ref{fig:four_viewings}. Unless stated otherwise, we take the camera at coordinate distance $d= 5 \sqrt Q$, with $d$ defined as in \eqref{eq:def_observer_position_inclination}.

\begin{table}[ht!]
\centering
\begin{tabular}{cccccc}
 Coord.\ & Topology & Inclination & along LOS & along screen\\
 \hline
 $x^2=0$ & 2 $S^2$'s& $i=0$ (Face on) &$/$ &  $J_{34}$ \\
  $x^2=0$ & 2 $S^2$'s& $i=\frac \pi2$ (Edge on)  &  $J_{34}$ &$/$\\
 \hline
  $x^4=0$ & $S^1 \times S^1$ & $i=0$ (Face on) & $/$ & $J_{12}$ \\
   $x^4=0$ & $S^1 \times S^1$ & $i=\frac \pi2$ (Edge on) &  $J_{12}$  &$/$\\

 \hline
\end{tabular}
\caption{The four main viewings we discuss in this paper, organized in two slices and two inclinations. For each we give the fixed Cartesian coordinate defining a 3D hypersurface, the topology of the two-dimensional slice of the horizon within this hypersurface, the inclination and whether the remaining angular momentum in the 3D hypersurface is along the line of sight (LOS) or the screen.}
\label{tab:four_viewings}
\end{table}

\subsubsection{Disconnected Two-sphere Topology}
First we fix $x^2=0$ to obtain the 2D slicing of the horizon with disconnected two-spheres (left 2 columns of figure \ref{fig:four_viewings}). As one might expect, the images show a striking resemblance to earlier visualizations of binary black holes \cite{Bohn:2014xxa} and we recover the main features. When viewed face on (inclination $i=0$), we clearly see the shape of the two disconnected parts of the horizon. At the outer edges, we also recognize the distinct `eyebrows', the flattened image of one part of the horizon from lensing around the other disconnected part \cite{Nitta:2011in,Yumoto:2012kz}. When imaging this slicing edge on (inclination $i = \pi/2$), the $S^2$'s and the camera align. The shadow of the farthest $S^2$ appears as a ring around the central shadow of the closer object. Such an image is similar to the visualization of an in-spiraling binary system shown in \cite{Bohn:2014xxa}. 

\begin{figure}[ht!]
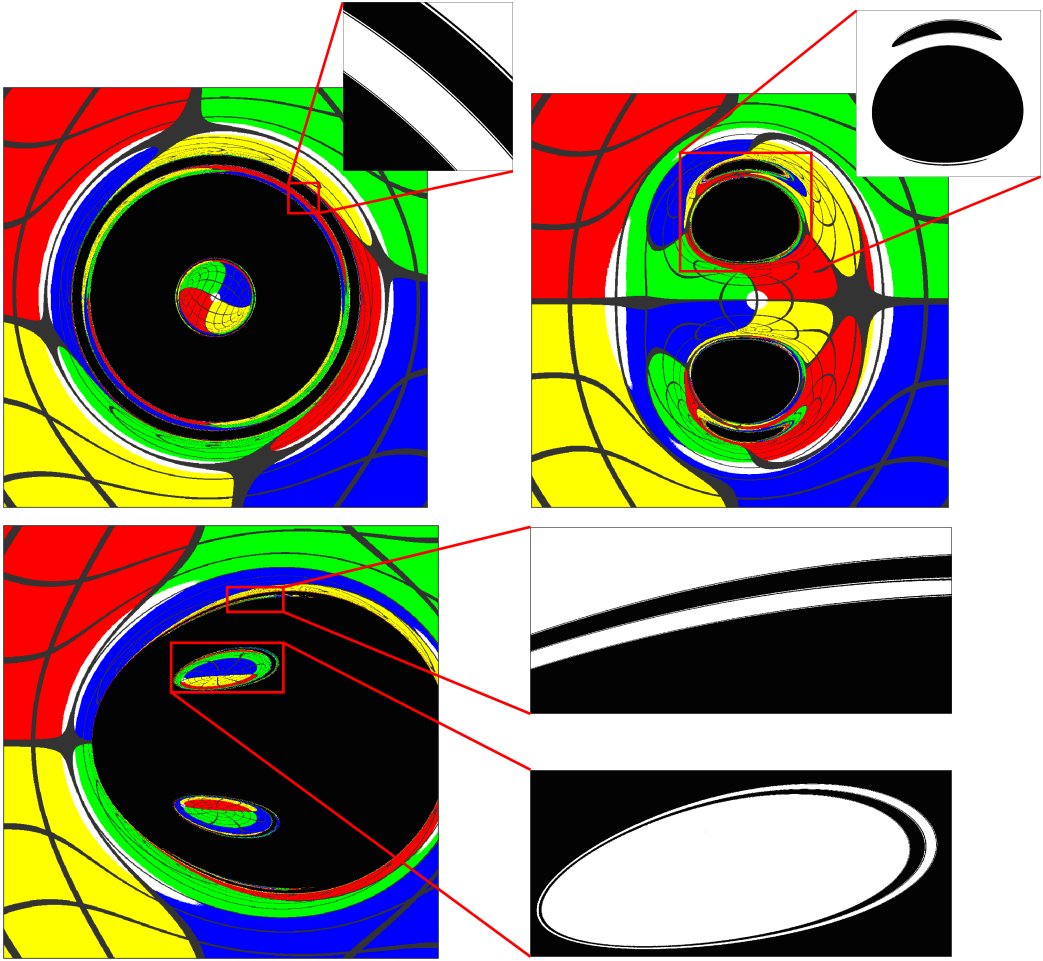

	\centering
	\resizebox {.45\linewidth} {!} {\input{./Images/Tikz/zoom_1}}
	\resizebox {.45\linewidth} {!} {\input{./Images/Tikz/zoom_2}}\\[2mm]
	\centering
	\resizebox {.91\linewidth} {!} {\input{./Images/Tikz/zoom_3}}
	\caption{High-resolution zoom picture of a portion of the shadow of black ring 4 viewed as in figure \ref{fig:four_viewings}. We clearly see the repetition of multiple annuli, eyebrows and structure inside the holes.}
	\label{fig:fractals}
\end{figure}

To investigate the appearance of chaotic effects due to lack of integrability, we zoomed in on one of the images, see figure \ref{fig:fractals}. Binary black hole shadows are known to exhibit fractal structure \cite{Bohn:2014xxa,Shipley:2016omi} and we confirm similar behaviour for black ring images. The black ring shadow shows a repeated structure of multiple eyebrows or concentric annuli, see figure \ref{fig:fractals}. Although a more detailed study is outside the scope of this work, we think it would be a worthwhile to investigate this further in terms of distinct loci in spacetime and Lyapunov exponents according the lines of \cite{Grover:2017mhm}.

We leave a more detailed comparison to known examples of two-center black holes to future work. For the supersymmetric rings we discuss in this paper, such a comparison would be most natural with extremal charged four-dimensional solutions, such as the two-center Papapetrou-Majumdar solution discussed in \cite{Yumoto:2012kz}. The extension to uncharged black rings would be interesting as well. 

\subsubsection{Toroidal Topology}
Next we fix $x^4 = 0$. The 2D slice of the black ring horizon has the topology of $S^1 \times S^1$, a two-dimensional torus. This is the first example of an image of a toroidal black shadow and its effects on the surrounding spacetime. 

The two inclinations we choose show a number of new effects. When viewed face on (third column of figure \ref{fig:four_viewings}), one can clearly distinguish two concentric black annuli: a central annulus with a black halo around it. Increasing precision reveals again a fractal-like looking structure of ever thinner concentric annuli as one moves away from the center of the image (figure \ref{fig:fractals}). The innermost ring indicates geodesics hitting the ring directly, the secondary ring those geodesics that have one turning point and so on. The thickness of the inner ring depends on the ring parameters: for large values of $L/q$, the ring radius greatly exceeds the radius of the $S^2$ cross-section and we naturally observe thinner rings with thinner annuli as a consequence.

When viewed edge on (fourth column of \ref{fig:four_viewings}), the image becomes most interesting. The shadow has three main features: an elongated shape, two holes, and a clear left-right asymmetry not of the D-shape kind known from earlier visualizations of rotating black objects:
\begin{itemize}
 \item The elongated shape reveals the shape of the ring. For larger values of $L/q$, the ring becomes thinner and the image becomes more elongated.
 \item The holes in the image arise from geodesics shooting through the ring from either above or below and escaping through infinity. By symmetry of the problem, the holes are symmetrically placed with respect to the horizontal axis and the holes intersect the vertical axis, as geodesics going through the ring and the origin of our coordinate system are always mapped to the vertical. We clearly distinguish two large holes near the horizontal axis and long thin holes near the shadow edge, with repeated shadow structure inside the holes (figure \ref{fig:fractals})\footnote{For completeness we note that in Fig 4a in \cite{Wang:2018aa} a shadow of a Manko-Novikov solution is reported displaying two holes.}.
 \item Both the holes and the shape of the shadow have a left-right asymmetry.  For generic values of angular momenta, the shadow edges left and right are smoothly curved, with a smaller radius of curvature on the left part of the figure. The asymmetry itself is caused by the rotation along the ring plane ($J_{12}$, along line of sight). The left-hand side of the image shows a part of the ring that is rotating towards the camera, which causes a contraction effect in the lensing, while the opposite is true for the right-hand side. The same effect happens for the holes. The angular momentum $J_{34}$ is only indirectly visible in the image: for fixed values of the $J_{12}$, it influences the size of the ring and the shape of the shadow.
\end{itemize}
All three qualitative features are related to the topology and affected by the ratio of the two angular momenta as we discuss now.

\subsubsection{Effects of Angular Momentum}\label{ss:compare_blackhole}

\begin{figure}[ht!]
\centering
\includegraphics[width=0.3\linewidth]{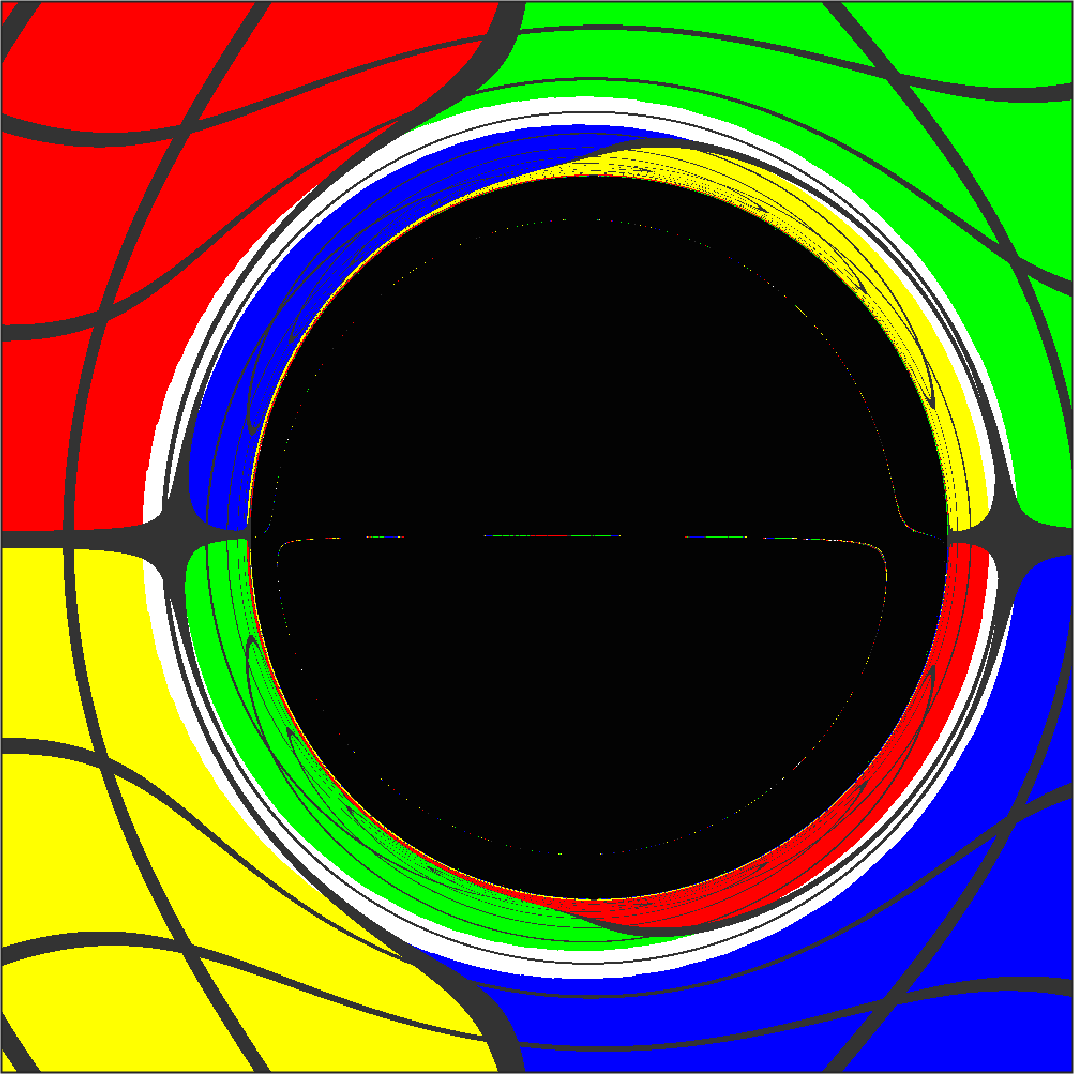}\includegraphics[width=0.3\linewidth]{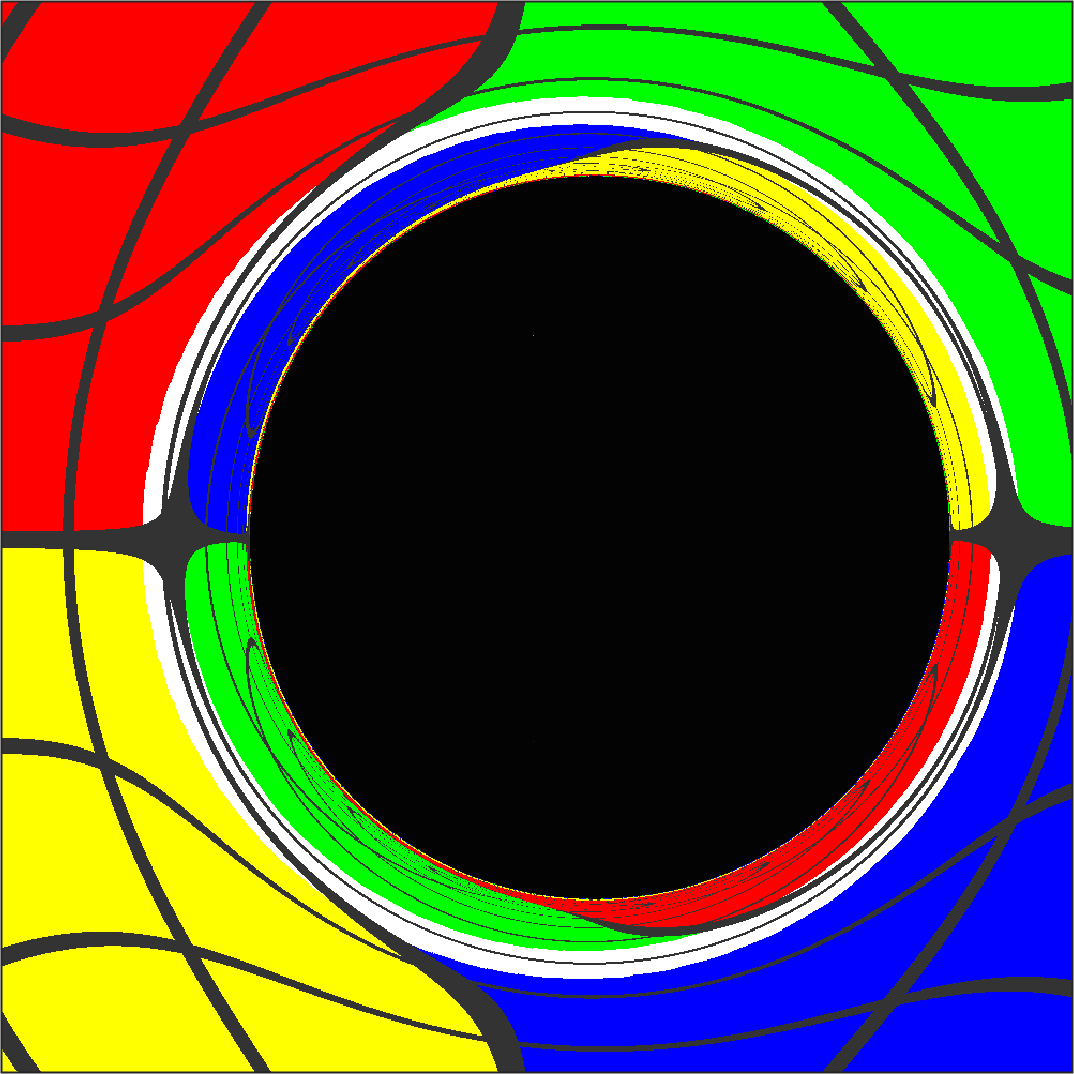}\\
\includegraphics[width=0.3\linewidth]{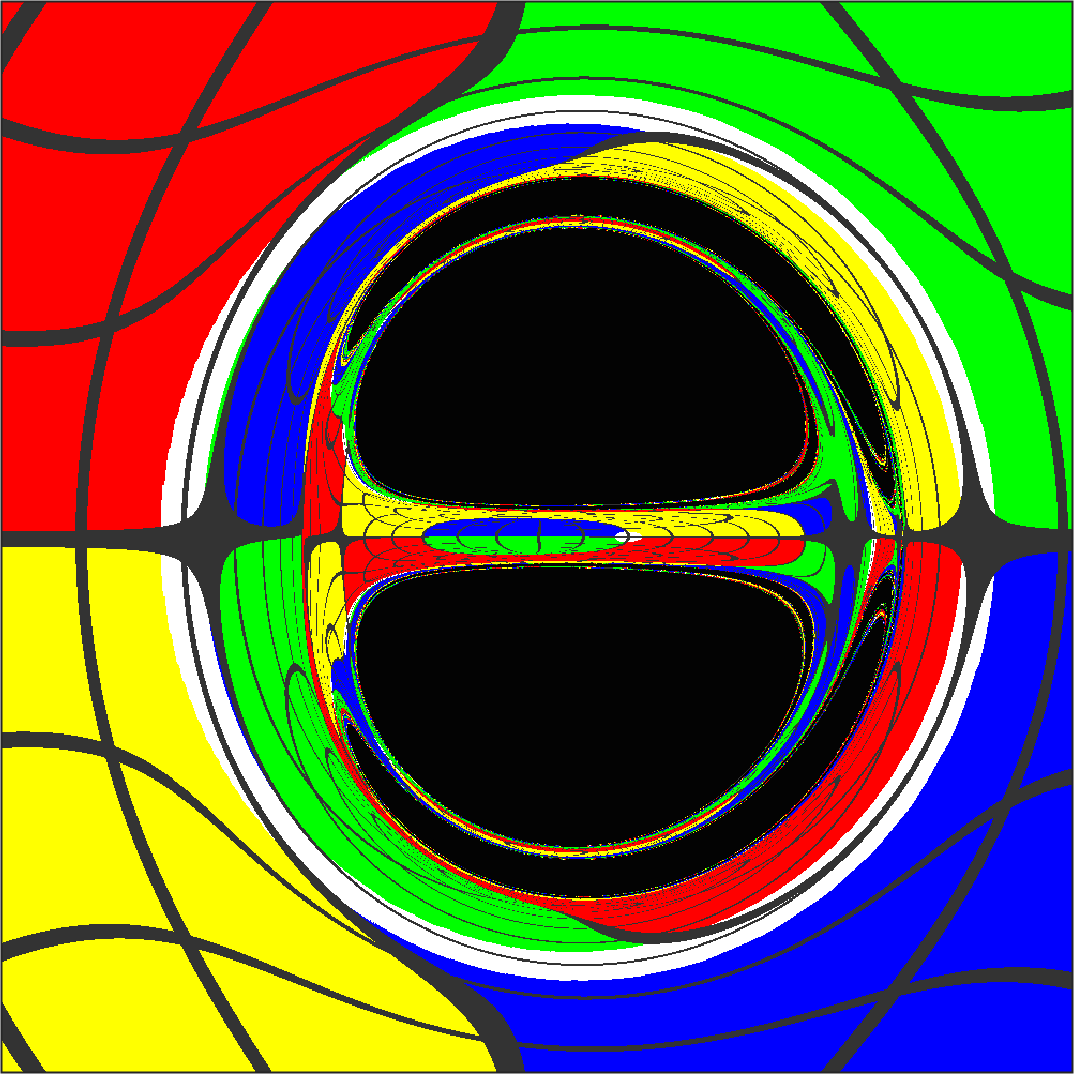}\includegraphics[width=0.3\linewidth]{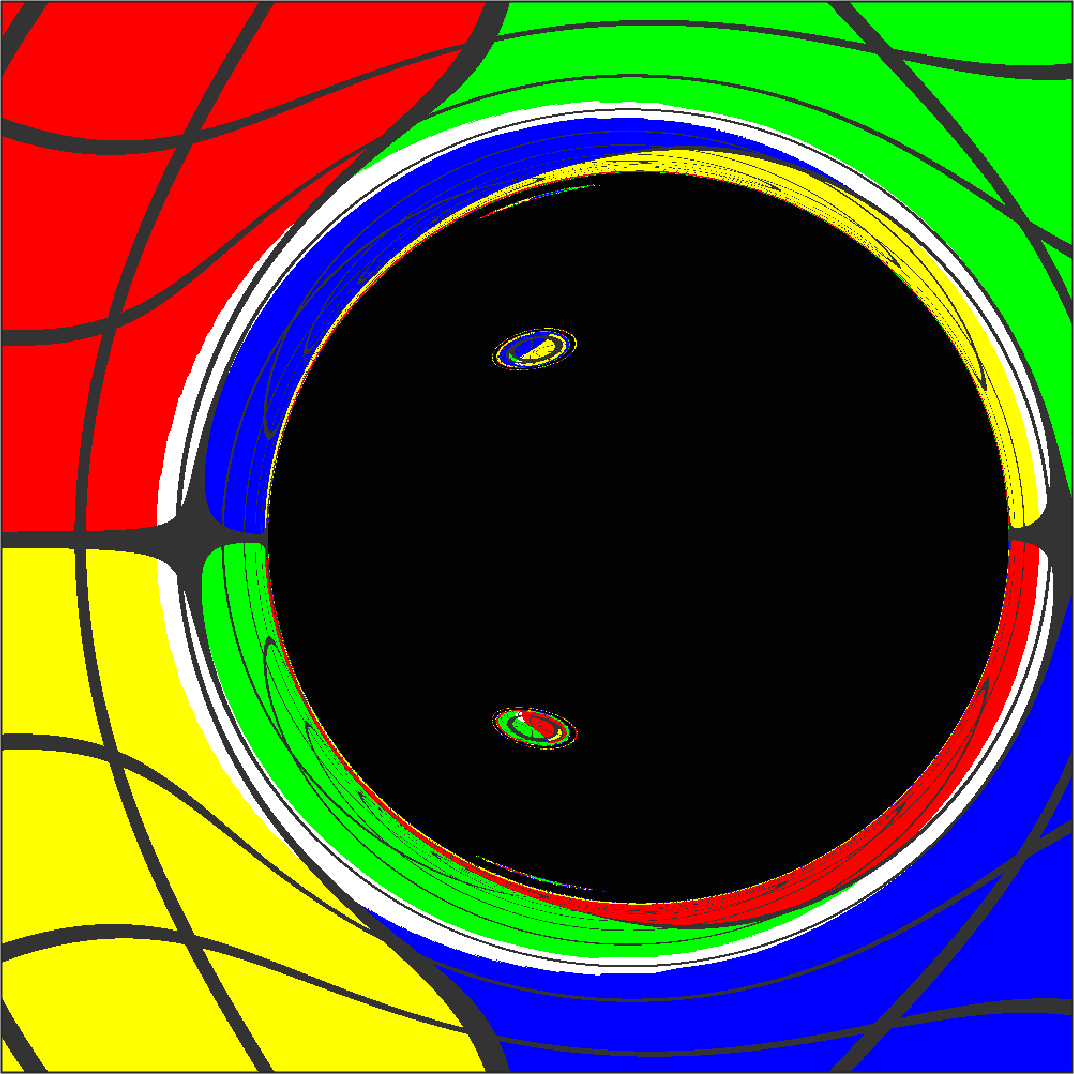}\\
\includegraphics[width=0.3\linewidth]{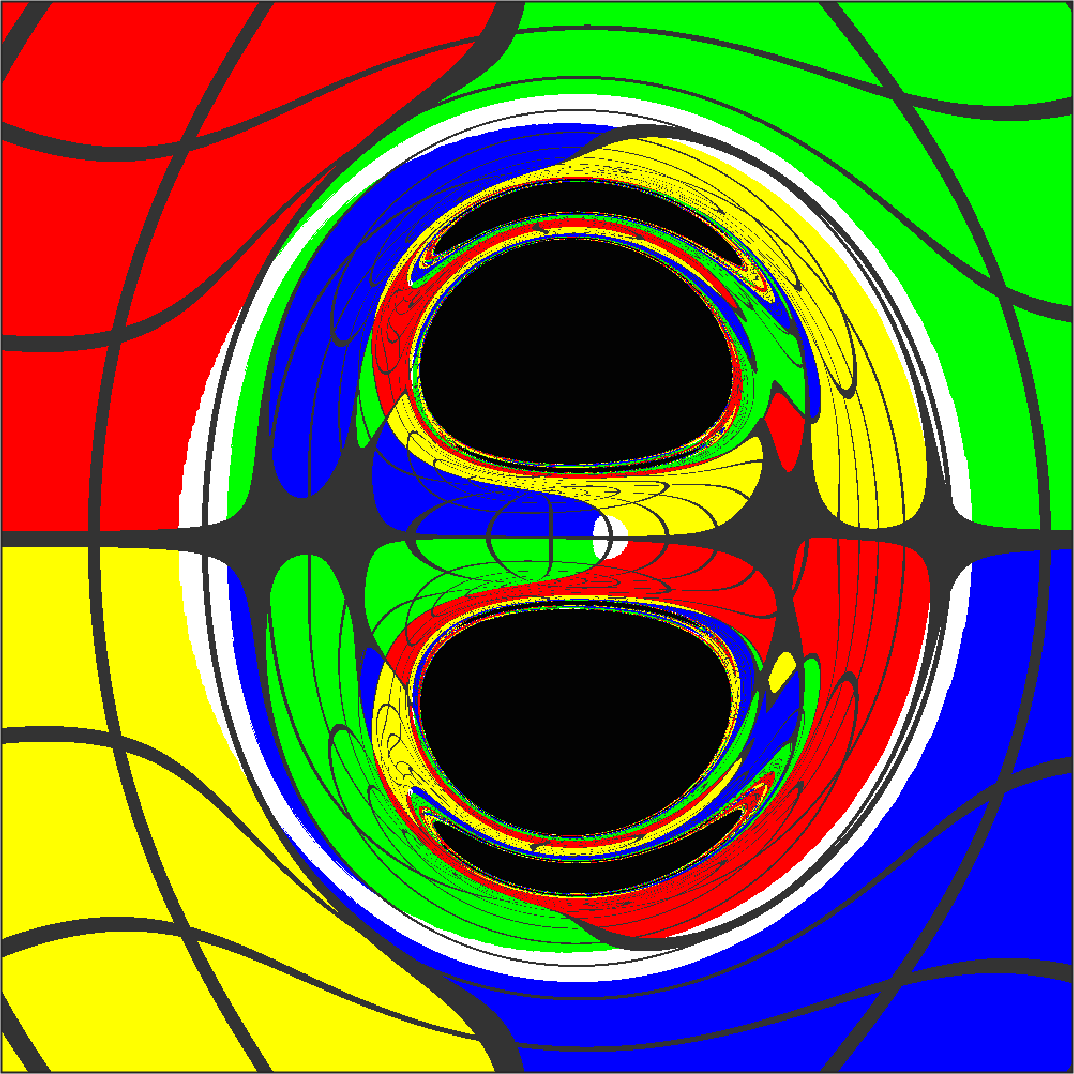}\includegraphics[width=0.3\linewidth]{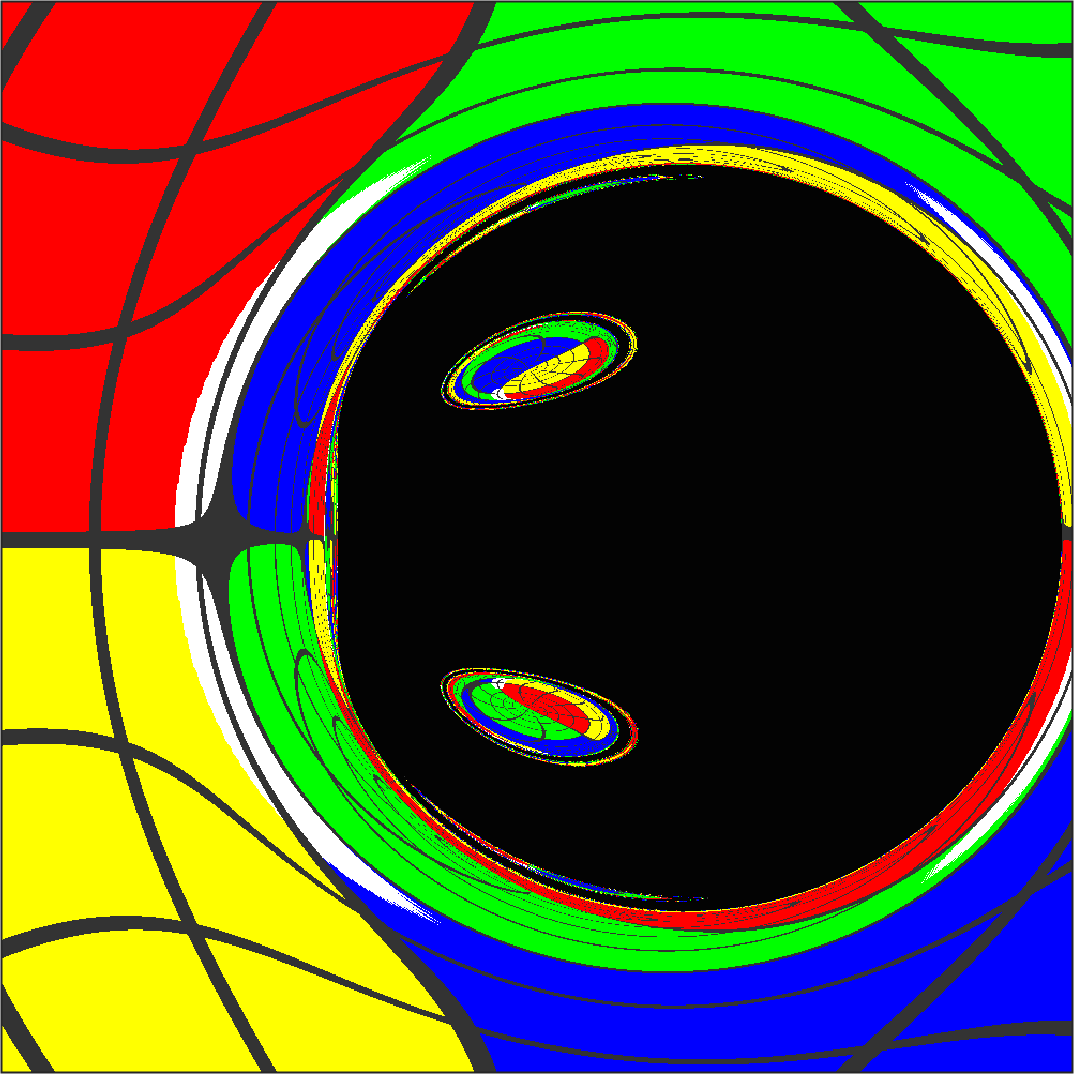} 
\caption{Images of black rings for constant $J_{34}=\frac{11}{16} Q^{3/2}$ but varying $J_{12}$. Rows represent black ring solutions 1,2,3 (top to bottom) in the phase diagram of figure \ref{fig:phasediagram1}. The bottom row is the extremal ring. Viewing angles and slicings correspond to left and rightmost columns of figure \ref{fig:four_viewings}. }
\label{fig:angular_momenta_1}
\end{figure}

We explore the phase space of allowed angular momenta for the black rings and compare to black holes with the same value of one of the two angular momenta. Recall that the $J_{12}$ is along the ring horizon $S^1$, while $J_{34}$ is orthogonal. figure \ref{fig:four_viewings} shows two different solutions with constant $J_{12}$, figure \ref{fig:angular_momenta_1} shows three different solutions with constant $J_{34}$. 

When $J_{12} \approx J_{34}$ (left column in figure \ref{fig:angular_momenta_1}), black ring parameters are close the the black hole and unsurprisingly the image is almost indistinguishable from a black hole's. 
In general, however, images of black rings differ quite a lot from the black hole image, and there is a lot of variation in size of the black ring compared to the black hole. 

\begin{figure}[ht!]
	\begin{center}
		\includegraphics[width=0.3\linewidth]{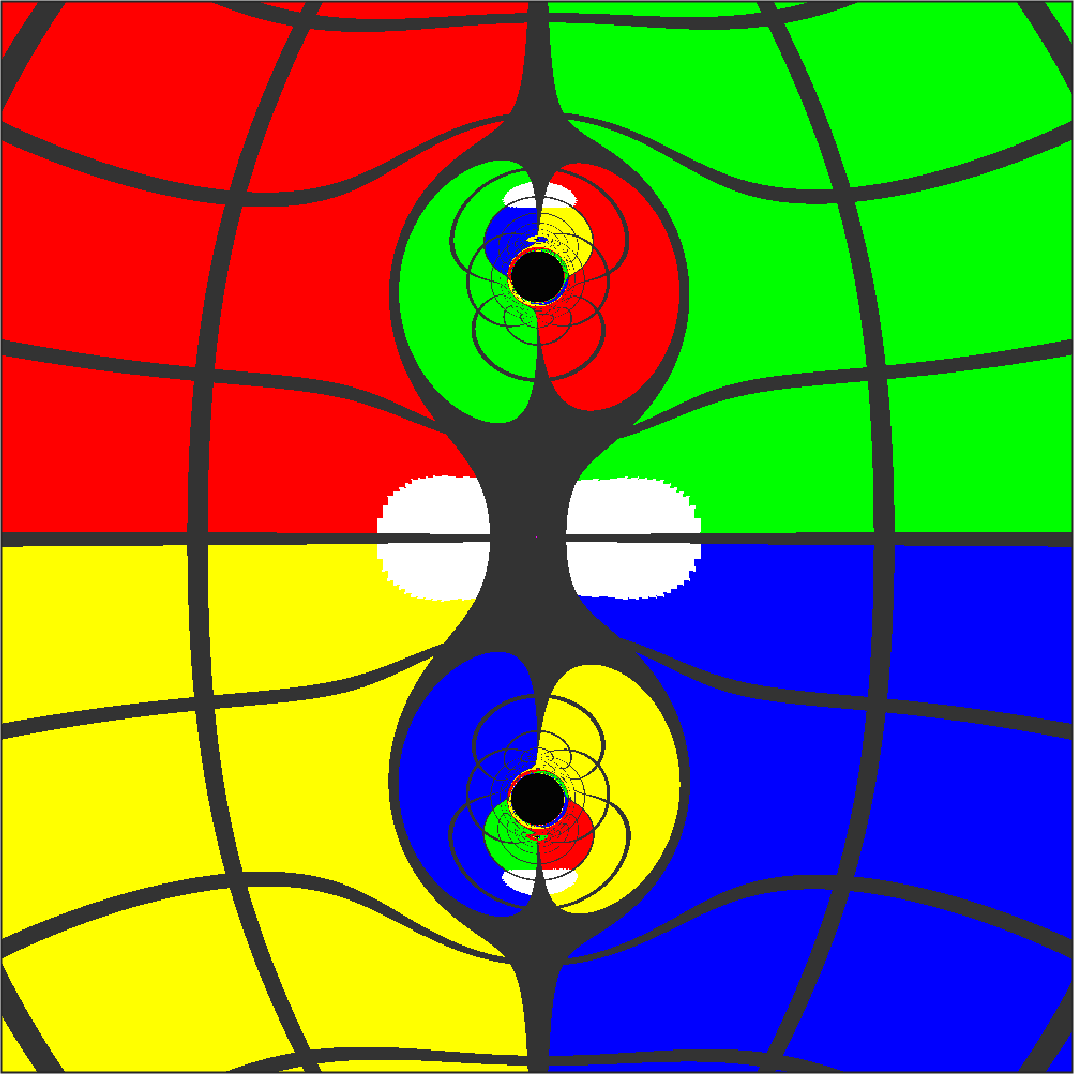} 
		\includegraphics[width=0.4625\linewidth]{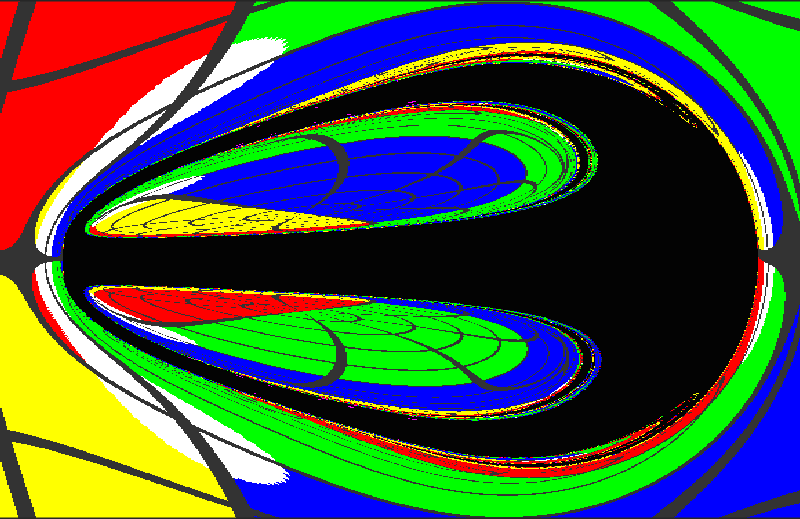}
	\end{center}
	\caption{Images of the extremal thin black ring solution 6 in the phase diagram figure \ref{fig:phasediagram1}, with $J_{12}/J_{34}= 3889/214 \approx 18.17 Q^{3/2}$ and $J_{12} = \frac {3889}{864} Q^{3/2}$, taken from coordinate distance $d=10\sqrt Q$.}
	\label{fig:angular_momenta_2}
\end{figure}

Having either one angular momentum along the line of sight, gives the clearest distinction of the black hole image, as it reveals either the disconnected or the elongated shape of the shadow, see figure \ref{fig:angular_momenta_1}. The bottom row in that figure shows an extremal spinning black ring, with maximal $J_{12}$ for given $J_{34}$. The effect of rotation on the surrounding spacetime is visible in all images, but the effect on the shadow is as announced only visible when the angular momentum along the line of sight, that is, for the viewing angles of figure \ref{fig:angular_momenta_1}. The edge on view of the disconnected slicing (bottom left) does not shows the typical D-like shape in the individual shadows of the disconnected parts of the images, but we can see this effect still on the image outside the shadow, which shows resemblance to the extremal black hole image of figure \ref{fig:BMPV_numeric}. The edge-on view of the toroidal slice (bottom right) is interesting: the shadow acquires the typical D-like shape and shift off-center towards the right, as known for near-extremal Kerr. But still we see the holes characteristic to black ring solutions.

Finally we want to point out that extremal rings with larger ratios of the angular momenta resemble much less the black hole image, see figure \ref{fig:angular_momenta_2}. Such rings are very thin and their shadows remain elongated; the flattening of the shadow is present only at the very tip of the shadow.

\section{Outlook}\label{sec:conclusions}

We have developed the numerical tools to explore multi-center black hole like solutions of string theory with null geodesics. As an application, we discussed images and shadows of supersymmetric black holes and black rings in five dimensions and we also showed how they give insight in four-dimensional physics. In this respect it is worth noting that  the visualization of black rings in higher dimensions offers a novel and complementary view on binary systems in four dimensions. This may prove important since the imaging of binaries in $3+1$ dimensions has some technical limitations arising from the fact that one is mostly restricted to working with numerical solutions \cite{Bohn:2014xxa} or solutions with conical singularities, such as the double Schwarzschild solution and double-Kerr solutions \cite{Cunha:2018gql,Cunha:2018cof}. By contrast black rings are analytically known and non-singular.
In this context it would be very interesting to adapt the quasi-static procedure of \cite{Cunha:2018cof} and consider images of rotating black rings. We leave this to future work.

There are two further directions in which our calculations should be extended. 
First, with our ray-tracing and visualization technique in place, it can be immediately applied to more intricate black hole like solutions. We are currently working on imaging solutions with more than two centers in the wide class of five-dimensional multi-center solutions, such as multi-black holes, multi-rings, and horizon-less microstate geometries \cite{Hertog:inprep1}. Ultimately, we deem it important to extend the method developed here to more generic spacetimes not restricted by supersymmetry.

Second we want to emphasize that 2D images using backwards ray-tracing are merely scratching the surface. There is a much broader set of analytic and numerical tools at our disposal to explore and study the spacetime structure in the neighbourhood of black objects. A very useful one, that can be obtained with similar integration methods, is the geodesic deviation near compact objects. As was noted in \cite{Tyukov:2017uig,Bena:2018mpb} geodesic probes can reach extremely high tidal forces long before they reach the region where the geometry differs significantly from the corresponding black hole.  We aim to use our code to check this explicitly for five-dimensional multi-center solutions. Other important extensions include using massive probes and the study of the effect of an environment around compact objects. 

More generally speaking, studies along these lines (see also e.g. \cite{Giddings:2016btb}) work towards bridging the gap between observations and string theory or quantum gravity. Imaging the shadow of resolvable supermassive black holes using upcoming results of the Event Horizon Telescope, but also the details of the ringdown encoded in gravitational wave signals of compact binaries, are largely determined by the nature of the light ring, the unstable photon orbit outside the black hole. Understanding the geodesic structure near and inside the light ring of alternative compact objects is thus key to flesh out any possible deviations from the GR paradigm; imaging objects constitutes an important step in that direction.

%
%

\section*{Acknowledgments}
We thank Fabio Bacchini, Vitor Cardoso, Pedro Cunha, Roberto Emparan, Carlos Herdeiro, Stefanos Katmadas, Sera Markoff, Marina Martinez, Ben Niehoff, Bart Ripperda and Erik Verlinde for discussions; Fabio and Bart Ripperda also for input from a related collaboration and Perdo Cunha and Carlos Herdeiro for useful feedback on the manuscript. The authors are partially supported by the National  Science  Foundation  of Belgium (FWO) grant G.001.12, the European Research Council grant no.\ ERC-2013-CoG 616732 HoloQosmos, the KU Leuven C1 grant ZKD1118 C16/16/005, the FWO Odysseus grant G0H9318N and the COST actions CA16104 \emph{GWVerse} and MP1210 The String Theory Universe.

\appendix

\section{Multi-center Solutions in 5D}\label{sec:metrics}

In this section we give the larger class of metrics we study  in one fixed notation, as different conventions have been used throughout the literature. We mostly follow \cite{Bena:2007kg}.

We focus on supersymmetric, stationary, asymptotically flat solutions of 5D supergravity couple to an arbitrary number of vector multiplets with a timelike Killing vector. This class of solutions is written as a fibration of time over a four-dimensional space space \cite{Gauntlett:2002nw,Gauntlett:2003fk,Gutowski:2004yv}
\begin{equation}
 ds^2 = -Z^{-2} (dt + k)^2 + Z ds_4^2\,.\label{eq:GHmetric}
\end{equation}
Supersymmetry requires the four-dimensional space to be hyperk\"ahler. 
We restrict to the large subset of metrics for which the four-dimensional base space is a Gibbons-Hawking space.  Those multi-center solutions were derived in five dimensions in \cite{Bena:2004de,Gauntlett:2004qy,Elvang:2004ds} and reviewed in \cite{Bena:2007kg}. The generic solution after reduction over the Gibbons-Hawking fibre originally given in \cite{Denef:2000nb,Bates:2003vx}. For concreteness, we present the code and the solutions in the `STU model' that arises in compactification of eleven-dimensional M-theory on $\mathbb{T}^6/\mathbb{Z}_2\times \mathbb{Z}_2$(2 vector multiplets). 

In section \ref{ssec:multi-center} we only present the solution of the metric, not of the gauge fields nor the scalar fields, as those are not needed for our purposes. In section \ref{ssec:explicit} we write out the metric for an axisymmetric but otherwise arbitrary solution in the class. We recall how the black hole and black ring metrics we use are recovered in section \ref{ssec:BH_BR}.

\subsection{Multi-center Solutions}\label{ssec:multi-center}

The family of metrics of three--charge solutions with a Gibbons--Hawking base space is determined by eight harmonic functions on flat $\mathbb{R}^3$, with $N$ centers:
\begin{IEEEeqnarray}{rClCrCl}
V &=&  v_{0} + \sum_{i=1}^N \frac{v_{i}}{r_{i}} & \qquad & K^{I} &=& k_{0}^{I} + \sum_{i=1}^N \frac{k^{I}_{i}}{r_{i}}\\
L_{I} &=& l_{0,I} + \sum_{i=1}^N \frac{\ell_{I,i}}{r_{i}} & \qquad & M &=& m_{0} + \sum_{i=1}^N \frac{m_{i}}{r_{i}},
\end{IEEEeqnarray}
where $r_{i} = \vert\vert \vec{r} - \vec{r}_{i} \vert\vert$, $\vec{r}_{i} \in \mathbb{R}^3$ are the locations of the centers which carry the harmonic charges and $I = 1,2,3$ is an index that runs over the gauge fields (2 from the vector multiplets, one from the graviphoton). Those harmonic functions enter the metric \eqref{eq:GHmetric} through intermediary functions $Z_I$, $\mu$ and one-forms $A,\omega$:
\begin{IEEEeqnarray}{rCl}
Z = (Z_1 Z_2 Z_3)^{1/3}\,,\qquad k =  \mu\left(\dif \psi+ A \right) + \omega\,,\qquad ds_4^2 = V^{-1}\left(\dif \psi+A\right)^2 + V \dif s_{3}^2,\nonumber
\end{IEEEeqnarray}
with the relations
\begin{IEEEeqnarray}{rCl}
	Z_{I} &=& L_{I} + \frac{1}{2}\vert \epsilon_{IJK} \vert \frac{K^{J}K^{K}}{V},\\
	\mu &=& \frac{K^{1}K^{2}K^{3}}{V^2} + \frac{K^{I} L_{I}}{2 V} + M,\\
	\star_{3} \dif A &=& \dif V,\\
	\star_{3} \dif \omega &=& V \dif M - M \dif V
	+ \; \frac{1}{2} \left(K^{I} \dif L_{I} - L_{I} \dif K^{I}\right). \IEEEeqnarraynumspace
\end{IEEEeqnarray}
Note that we allow the four-dimensional Gibbons-Hawking metric to be `ambipolar', that is: it can flip signature from $++++$ to $----$ between regions, by allowing $ZV$ to switch sign at the same location. The boundary region is known as an `evanescent ergosurface' \cite{Bena:2007kg,Gibbons:2013tqa}.

We get flat 5D asymptotics by setting $v_{0} =  k_{0}^{I} = 0$, $\ell_{0,I}  = \sum_{i=1}^N v_{i} = 1$ and $ m_{0} = -\frac{1}{2}\sum_{I,i}k^{I}_{i}$. Then the ADM mass and electric charges are
\begin{IEEEeqnarray}{rCl}
M &=& \frac{\pi}{4G_{5}}\left(Q_{1} + Q_{2} + Q_{3}\right),\\
Q_{I} &=& 4\sum_{i=1}^N\left(\ell_{I,i} + \frac{1}{2}\vert \epsilon_{IJK} \vert k^{J}_{i}k^{K}_{i}\right).
\end{IEEEeqnarray}
The angular momenta can be obtained from a Komar integral, see for instance \cite{Emparan:2008eg} for the normalization we use. We only give the full expression angular momenta below for axisymmetric solutions.

Generically these metrics will have CTC's, a necessary requirement to avoid these are the bubble equations:
\begin{IEEEeqnarray}{rCl}
	m_{0}v_{j} + \sum_{I}\frac{k^{I}_{j}}{2} &=& \sum_{i \neq j} \frac{v_{i}m_{j} - m_{i}v_{j} + \frac{1}{2} \left( k^{I}_{i}\ell_{I,j}  -  \ell_{I,i}k^{I}_{j}\right)}{\vert\vert \vec r_{i}-\vec r_{j} \vert\vert}, \qquad \forall j.\label{eq:bubble}
\end{IEEEeqnarray}
The bubble equations impose $N-1$ independent relations on the constants determining the solution

\subsection{Axisymmetric Solutions}\label{ssec:explicit}

If we assume that all centers are on on a line along the $z$--axis of the 3D space we can solve for the one-forms explicitly:
\begin{IEEEeqnarray}{rCl}
	A &\equiv& A_{\phi} \dif \phi=  \left( -1 + \sum_{i} v_{i}g_{i}\right) \dif \phi\\
	\omega &\equiv& \omega_{\phi} \dif \phi=\left(\frac{1}{2}\sum_{i,j} u_{i,j} f_{i,j} - \sum_{j} t_{j} g_{j}\right) \dif \phi,
\end{IEEEeqnarray}
with:
\begin{IEEEeqnarray}{rCl}
	g_{i} &=& \frac{r\cos \theta - z_{i}}{r_{i}}\\
	f_{i,j} &=& 
	\begin{cases}
		\!\begin{aligned}
			& \frac{r^2-\left(z_{i}+z_{j} - r_{i} + r_{j} \right)r\cos \theta}{(z_{i}-z_{j})r_{i}r_{j}}\\
			&  + \frac{z_{i}z_{j}  + z_{i}r_{j} - z_{j}r_{i} - r_{i}r_{j}}{(z_{i}-z_{j})r_{i}r_{j}}
		\end{aligned}   & \text{ when } i < j\\
		- f_{j,i} & \text{ when } j < i
	\end{cases}\\
	u_{i,j} &=& v_{i}m_{j} - m_{i}v_{j} + \frac{1}{2} \left( k^{I}_{i}\ell_{I,j}  -  \ell_{I,i}k^{I}_{j} \right)\\
	t_{j} &=& c_{m}v_{j} + \sum_{I}\frac{k^{I}_{j}}{2} - \sum_{i}\frac{u_{i,j}}{\vert z_{i}-z_{j} \vert},
\end{IEEEeqnarray}
where the $z_{i}$ are the locations of the centers along the $z$--axis. The functions $g_{i}$ in $\omega$ lead to  Dirac-Misner strings, to prevent CTC's we need to impose the bubble equations \eqref{eq:bubble}.

The angular momenta take the form
 \begin{IEEEeqnarray}{rCl}
J_{12} &=& \frac{2\pi}{G_{5}}\left[\sum_{i}m_{i}  + \frac{1}{2}\sum_{i,j}l_{i,I}k_{j}^{I} + \sum_{i,j,h}k_{i}^{1}k_{j}^{2}k_{h}^{3} + \frac{1}{2}\sum_{i,I}z_{i}\left(k_{i}^{I}-v_{i}\sum_{j}k_{j}^{I}\right)\right]\nonumber\\
J_{34} &=&  J_{12} - \frac{2\pi}{G_{5}}\left[\frac{1}{2}\sum_{i,I}z_{i}\left(k_{i}^{I}-v_{i}\sum_{j}k_{j}^{I}\right)\right].
\end{IEEEeqnarray}

\subsection{Supersymmetric Black Holes and Black Rings}\label{ssec:BH_BR}

The general supersymmetric black hole solution and black ring solutions in this class are the three-charge generalizations \cite{Breckenridge:1996sn,Elvang:2004ds} of the BMPV black hole \cite{Breckenridge:1996is} and supersymmetric black ring \cite{Elvang:2004rt} we imaged in this paper. We recall how those fit into the general class we give above.

\subsubsection{Three-charge Black Holes}

The three-charge black hole \cite{Breckenridge:1996sn} has a single center located at the origin of $\mathbb{R}^3$ with the Gibbons-Hawking charges: 
\begin{equation}
	v = 1  \quad  k^{I} = 0,\quad \ell_{I} = \frac{Q_{I}}{4}  \quad  m =  \frac J 8.
\end{equation}
This solution has three electric charges $Q_{I}$ and angular momenta $J_{12} = J_{34} = J$. The metric is determined by:
\begin{equation}
Z_{I} =1 + \frac{Q_{I}}{4r},  \qquad  \mu =\frac{J}{8 r},\qquad 
A = \left( 1 + \cos \theta\right) \dif \phi, \qquad \omega = 0.
\end{equation}
The black hole metric can be brought into a more standard form upon performing the coordinate transformation
$r = \frac{\rho^2 - Q}{4}, \tilde \theta = 2\theta - \pi, \phi_1 = \pi + \phi - \psi, \phi_2 = 2 \psi - 2\pi$:
\begin{IEEEeqnarray}{rCl}
 ds^2_4 &=& - f^{2} (dt + k)^2 + f^{-1} (d\rho^2 + \rho^2 (d\tilde \theta^2 + \sin^2 \theta  d\phi_1^2 + \cos^2 \theta d\phi_2^2))\,, 
\end{IEEEeqnarray}
with $f\equiv Z^{-1} = 1 - \frac Q \rho$ .

In this paper, we take the three charges to be equal  $Q_{1} = Q_{2} = Q_{3} \equiv Q$ to obtain the BMPV solution \cite{Breckenridge:1996is,Diemer:2013fza}.

\subsubsection{Three Charge Black Rings} 

We write the black ring of \cite{Elvang:2004ds}  in our conventions. We take two centers located at $\vec r_1 = \vec 0$ and $\vec r_2 = a \vec u_z$, with $\vec u_z$ a unit vector along the $z$-axis in $\mathbb{R}^3$. 
The harmonic functions are 
\begin{equation}
V = \frac 1 r\,,\qquad K^I = \frac {q^I}{2\Sigma}\,,\qquad L_I = \frac{\bar Q_I}{4 \Sigma}\,,\qquad M = \frac 1 4(q^1 + q^2 + q^3) \left(-1 + \frac{a} \Sigma\right)\,,  
\end{equation}
with 
\begin{equation}
 \Sigma\equiv ||\vec r_2 ||  =  \Sigma = \sqrt{r^2 + a^2 - 2 r a \cos \theta}\,,\qquad  \bar Q_I = Q_I - \frac 12  C_{IJK} q^J q^K\,,
\end{equation}
The relation to the ring parameter $R$ is  $a = R^2/4$. The functions appearing in the metric \eqref{eq:GHmetric} are
\begin{IEEEeqnarray}{rCl}
 Z_I &=&1+ \frac{\bar Q_I}{4 \Sigma } + \frac 12 \frac{C_{IJK} q^J q^K}{4 \Sigma ^2},\nonumber\\
 \omega &=&  \frac 1 4(q^1 + q^2 + q^3) \left(1 - \frac a\Sigma  - \frac r\Sigma \right)(\cos \theta -1) \dif \phi\,,\nonumber\\
  \mu &=& - \frac 1 4 (q^1 + q^2 + q^3) \left(1 - \frac a \Sigma - \frac r \Sigma\right) + \frac r {16 \Sigma^2}\left(  q^I \bar Q_I+ 2  q^1 q^2 q^3\frac r \Sigma \right)\,,\\
  A &=& (1 + \cos \theta)d\phi\,.
\end{IEEEeqnarray} 

The spacetime has ADM mass, electric charges and angular momenta:
\begin{IEEEeqnarray}{rCl}
M &=& \frac \pi {4 G_5} (Q_1 + Q_2 + Q_3)\\
Q_{I} &=& \bar{Q}_{I} + \frac{1}{2}\vert \epsilon_{IJK} \vert q^{J}q^{K}\\
J_{12} 
&=& \frac{\pi}{4 G_{5}}\left[\frac{1}{2}\left(q_{I}Q^{I} - q_{1}q_{2}q_{3}\right) + 4a\left(q_{1} + q_{2} + q_{3}\right)\right]\\
J_{34} 
&=& \frac{\pi}{4 G_{5}}\left[\frac{1}{2}\left(q_{I}Q^{I} - q_{1}q_{2}q_{3}\right)\right].
\end{IEEEeqnarray}
This solution also has dipole charges equal to $ \frac{q^{I}}{2 \pi} $. The bubble equations are satisfied with this choice of centers. Plugging this second configuration into the metric yields the extremal three-charge black ring in a form similar to equation (2.6) in \cite{Bena:2005ni}.

In this paper, we present the black ring with $Q \equiv Q_1 = Q_2 = Q_3$ and $q \equiv q^1 = q^2 =q^3$.

\section{Geodesics}\label{sec:geodesics}

In this appendix we give the geodesic equations for the general axisymmetric multi-center solutions and give the initial conditions for the setup described in \ref{sec:tech}. We have three isometries in the spacetimes with all centers along the $z$ axis, these lead to the following constants of motion for geodesics:
\begin{IEEEeqnarray}{rCl}
	E &=& \frac{1}{Z^2}\left[\frac{\dif t}{\dif \lambda} + \left(\mu A_{\phi} + \omega_{\phi}\right)\frac{\dif \phi}{\dif \lambda} + \mu \frac{\dif \psi}{\dif \lambda} \right]\\
	L_{\phi} &=& -\omega_{\phi} E + A_{\phi} L_{\psi}+ ZV \rho^2 \sin^2(\theta)\frac{\dif \phi}{\dif \lambda} \\
	L_{\psi} &=& - \mu E + \frac{A_{\phi} Z}{V}\frac{\dif \phi}{\dif \lambda}  + \frac{Z}{V} \frac{\dif \psi}{\dif \lambda}.
\end{IEEEeqnarray}
Those can be solved for the derivatives to get a set of first order equations:
\begin{IEEEeqnarray}{rCl}
	\frac{\dif t}{\dif \lambda} &=& Z^2E - \left(\mu A_{\phi} + \omega_{\phi}\right)\frac{\dif \phi}{\dif \lambda} - \mu \frac{\dif \psi}{\dif \lambda}\\
	\frac{\dif \phi}{\dif \lambda} &=& \frac{L_{\phi} - A_{\phi}L_{\psi} + E\omega_{\phi}}{VZ r^2 \sin^2(\theta)}\\
	\frac{\dif \psi}{\dif \lambda} &=& \frac{V\left(L_{\psi} + E\mu\right)}{Z} - A_{\phi}\frac{\dif \phi}{\dif \lambda}.
\end{IEEEeqnarray}
In general we cannot find first order equations for $r$ and $\theta$, we have to resort to the second order geodesic equation so we need the Christoffel symbols:
\begin{IEEEeqnarray}{rCl}
	\Gamma_{\nu \sigma}^{r} &=& \frac{2\delta_{\sigma}^{r}\delta_{\nu}^{r}\partial_{r}VZ + \left(\delta_{\sigma}^{r}\delta_{\nu}^{\theta} + \delta_{\sigma}^{\theta}\delta_{\nu}^{r}\right)\partial_{\theta}VZ - \partial_{r}g_{\nu \sigma}}{2VZ}\\
	\Gamma_{\nu \sigma}^{\theta} &=& \frac{2\delta_{\sigma}^{\theta}\delta_{\nu}^{\theta}\partial_{\theta}r^2VZ + \left(\delta_{\sigma}^{\theta}\delta_{\nu}^{r} + \delta_{\sigma}^{r}\delta_{\nu}^{\theta}\right)\partial_{r}r^2VZ - \partial_{\theta}g_{\nu \sigma}}{2r^2VZ}.
\end{IEEEeqnarray}

There is one more constant of motion:
\begin{equation}
 \delta = g_{\mu \nu}\frac{\dif x^{\mu}}{\dif \lambda}\frac{\dif x^{\nu}}{\dif \lambda}
\end{equation}
In principle this could be used to reduce one of the second order equations to a first order equation but since $\delta$ is quadratic in the derivatives this is not very practical. Instead we will use it to keep track of integration errors similar to what was done in \cite{Psaltis:2010ww}. We define a parameter:
\begin{IEEEeqnarray}{rCl}
	\xi &=& Z^3\left[\frac{\dif t}{\dif \lambda} + \left(\mu A_{\phi} + \omega_{\phi}\right) \frac{\dif \phi}{\dif \lambda} + \mu\frac{\dif \psi}{\dif \lambda}\right]^{-2}  \left[V^{-1}\left( A_{\phi}\frac{\dif \phi}{\dif \lambda} + \frac{\dif \psi}{\dif \lambda}\right)^2 \right. \nonumber\\
	&&  \qquad \left. + \; V \left(\left(\frac{\dif r}{\dif \lambda}\right)^2 + r^2\left(\frac{\dif \theta}{\dif \lambda}\right)^2 + r^2\sin^2(\theta)\left(\frac{\dif \phi}{\dif \lambda}\right)^2 \right)\right].\nonumber
\end{IEEEeqnarray}
The null condition $\delta = 0$ is then equivalent to $\xi = 1$.

We can simplify the geodesic equations by introducing a new parameter $\tau$ by $r^2 V Z \dif \tau = \dif \lambda$, we get:
\begin{IEEEeqnarray}{rCl}
	\frac{\dif t}{\dif \tau} &=& r^2 VZ^3E - \left(\mu A_{\phi} + \omega_{\phi}\right)\frac{\dif \phi}{\dif \tau} - \mu\frac{\dif \psi}{\dif \tau}, \nonumber\\
	\frac{\dif^2 r }{\dif \tau^2} &=& \frac{1}{V Z}\left[\frac{2 VZ}{r}\left(\frac{\dif r }{\dif \tau}\right)^2 + \frac{1}{2} \partial_{r}g_{\nu \sigma}\frac{\dif x^{\nu} }{\dif \tau}\frac{\dif x^{\sigma} }{\dif \tau}\right], \nonumber\\
	\frac{\dif^2 \theta }{\dif \tau^2} &=& \frac{1}{2 r^2 V Z}\partial_{\theta}g_{\nu \sigma}\frac{\dif x^{\nu} }{\dif \tau}\frac{\dif x^{\sigma} }{\dif \tau}, \label{eq:geod}\\
	\frac{\dif \phi}{\dif \tau} &=& \frac{L_{\phi} - A_{\phi}L_{\psi} + E\omega_{\phi}}{\sin^2\left(\theta\right)}, \nonumber\\
	\frac{\dif \psi}{\dif \tau} &=& r^2V^2\left(L_{\psi} + E\mu\right) - A_{\phi}\frac{\dif \phi}{\dif \tau},\nonumber
\end{IEEEeqnarray}
and for the error parameter:
\begin{IEEEeqnarray}{rCl}
	\xi &=& \left(r^4\sin^2\left(\theta\right) VZ^3E^2\right)^{-1}  \left[\sin^2\left(\theta\right)\left(\frac{\dif r}{\dif \tau}\right)^2 + r^2\sin^2\left(\theta\right)\left(\frac{\dif \theta}{\dif \tau}\right)^2 \right.  \\ 
	&& \qquad \left. + \; r^2\left(L_{\phi} - A_{\phi}L_{\psi} + E\omega_{\phi}\right)^2 + r^4\sin^2\left(\theta\right)V^2\left(L_{\psi} + E\mu\right)^2 \right]. \nonumber
\end{IEEEeqnarray}

With the setup for intial conditions described in \ref{sec:tech} we want to directly calculate the values of the constants of motion as functions of $(a,b,c)$ and $E$. First we use the null condition $\xi = 1$ to fix $k_{0}$, this yields:
\begin{IEEEeqnarray}{rCl}
	k_{0} &=& E \sqrt{\frac{Z}{V C}},
\end{IEEEeqnarray}
where:
\begin{equation}
	C = \frac{d^2}{4}\left(1 + \bar{v}_{z}^2 + \left(\cos(i) \bar{v}_{y} - \sin(i)\bar{v}_{w}\right)^2\right) + \frac{1}{d^2V^2}\left(\frac{A_{\phi}}{\sin i}\bar{v}_{y} - \frac{A_{\phi} -  2}{\cos i}\bar{v}_{w}\right)^2.
\end{equation}
Now we can just plug everything in and we have a full set of initial conditions and constants of motion:
\begin{IEEEeqnarray}{rCl}
	\left. \frac{\dif r}{\dif \tau}\right\vert_{\tau = 0} &=&  -E\left(\frac{d}{2}\right)^5 \sqrt{\frac{Z^3V}{C}}, \label{eq:init}\\
	\left. \frac{\dif \theta}{\dif \tau}\right\vert_{\tau = 0} &=&  \bar{v}_{z}E\left(\frac{d}{2}\right)^3 \sqrt{\frac{Z^3V}{C}}, \nonumber\\
	L_{\phi} &=& -\omega E + A L_{\psi} \nonumber\\
	&& - \; E \frac{d^3}{4}\sqrt{\frac{Z^3 V}{C}}\sin\left(i\right)\cos\left(i\right)\left(\cos\left(i\right) \bar{v}_{y} - \sin\left(i\right) \bar{v}_{w}\right), \nonumber\\
	L_{\psi} &=& - \mu E \nonumber\\
	&& - \; \frac{E}{ d}\sqrt{\frac{Z^3}{V^3C}}\left(\frac{A_{\phi}}{\sin\left(i\right)}\bar{v}_{y} - \frac{A_{\phi}-2}{\cos\left(i\right)}\bar{v}_{w}\right). \nonumber
\end{IEEEeqnarray}

\section{Exact BMPV Shadow}\label{sec:BMPV_exact}
%

We derive the first order geodesic equations for the 3-charge extension of the BMPV black hole \cite{Breckenridge:1996sn} using the Hamilton-Jacobi method and give the parameteric equation for the black hole shadow in terms of impact parameters and inclination angle.

\subsection{Geodesic Equations from Hamilton-Jacobi Method}

The Hamilton-Jacobi equations for geodesics is:
\begin{equation}
-\frac{\partial S}{\partial \lambda} = \frac{1}{2} g^{\mu \nu} \frac{\partial S}{\partial x^{\mu}}\frac{\partial S}{\partial x^{\nu}}\,.
\end{equation}
We use the ansatz:
\begin{equation}
S = \frac{\delta}{2} \lambda - Et + L_{\phi} \phi + L_{\psi} \psi + S_{r}(r) + S_{\theta}(\theta).
\end{equation}
The constants in this equation are interpreted as length of the geodesic $\delta$, energy $E$, angular momenta $L_\psi$ and $L_\phi$. Following the Hamilton-Jacobi procedure gives us the following set of equations:
\begin{IEEEeqnarray}{rCl}
	\frac{\dif t}{\dif \tau} &=& ErZ^3 - \frac{L_{\psi}J}{8r} - \frac{EJ^2}{64 r^2},\\
	\IEEEeqnarraymulticol{3}{l}{
		\left(\frac{\dif r}{\dif \tau}\right)^2 = R, \label{eq:radial_equation}
	}\\
	\IEEEeqnarraymulticol{3}{l}{
		\left(\frac{\dif \theta}{\dif \tau}\right)^2 = \Theta,
	}\\
	\frac{\dif \phi}{\dif \tau} &=& \frac{L_{\phi} - \left(1+\cos(\theta)\right)L_{\psi}}{\sin^2(\theta)},\\
	\frac{\dif \psi}{\dif \tau} &=& \frac{2 L_{\psi} - L_{\phi}}{1 - \cos(\theta)}  +  \frac{EJ}{8r},
\end{IEEEeqnarray}
with:
\begin{IEEEeqnarray}{rCl}
	\Theta &=& K_{c}-\frac{\left(L_{\phi} - \left(1+\cos(\theta)\right)L_{\psi}\right)^2}{\sin^2(\theta)} - L_{\psi}^2\,,\\
	R &=& r^3\left(E^2 Z^3 - \delta Z\right)-r^2K_{c} - \frac{E L_{\psi} J}{4}r - \frac{E^2 J^2}{64},
\end{IEEEeqnarray}
and where $K_{c}$ is the equivalent of the `Carter constant', an additional constant of motion given by:
\begin{IEEEeqnarray}{rCl}
	K_{c} &=& \frac{\left(L_{\phi} - \left(1+\cos(\theta)\right)L_{\psi} \right)^2}{\sin^2(\theta)} + L_{\psi}^2 + \left(\frac{\dif \theta}{\dif \tau}\right)^2.
\end{IEEEeqnarray}
Note that $\tau$ is related to the affine parameter $\lambda$ by $\dif \lambda = rZ\dif \tau$.

\subsection{Black Hole Shadow}

Only geodesics with impact parameters such that \eqref{eq:radial_equation} has a turning point escape to infinity; the others fall into the black hole. Hence the edge of the shadow corresponds to the family of null geodesics defined by ${\dif r}/{\dif \tau} = {\dif^2 r}/{\dif \tau^2} = 0$, or equivalently $R(r_*) = R'(r_*) = 0$. For null geodeiscs $\delta =0$ and we find that $R$ is given by:
\begin{IEEEeqnarray}{rCl}
	\frac{R}{E^2} &=& r^3 + \left(\frac{Q_{1} + Q_{2} + Q_{3}}{4} - \tilde{K}_{c}\right)r^2 + \frac{\left(Q_{1}Q_{2}Q_{3}- J^2\right)}{64} \nonumber\\
	&& + \; \left(\frac{Q_{1}Q_{2} + Q_{1}Q_{3} + Q_{2}Q_{3}}{16} - \frac{b_{\psi}J}{4}\right)r.
\end{IEEEeqnarray}
We have introduced the impact parameters
\begin{equation}
 b_\phi = \frac{L\phi}{E}\,,\qquad b_\phi = \frac{L\phi}{E}\,,\qquad \tilde K_c = \frac{K_c}{E}\,,
\end{equation}
such that $E$ drops out of the equations. 

We want to rework the equations $R(r_*) = R'(r_*) = 0$ to obtain the parametric dependence of the shadow edge on the impact parameters and the inclination angle of the observer.  Since for $J=0$ the impact parameter $b_{\psi}$ drops out of the equation, we have to distinguish between zero and non-zero rotation.


\subsubsection{Rotating Black Hole \texorpdfstring{($J\neq 0$)}{}}
Since $R$ is a polynomial, the condition $R(r_*) = R'(r_*) =0$ says that $R$ has a double root.
The solution for the double root condition is given by:
\begin{IEEEeqnarray}{rCl}
	\tilde{K}_{c}\left(r_{*}\right) &=& 2r_{*} + \frac{Q_{1} + Q_{2} + Q_{3}}{4} - \frac{Q_{1}Q_{2}Q_{3}- J^2}{64r_{*}^2},\\
	b_{\psi}\left(r_{*}\right) &=& -\frac {32r_{*}^3 - 2 r_{*}\left(Q_{1}Q_{2} + Q_{1}Q_{3} + Q_{2}Q_{3}\right) - \left(Q_{1}Q_{2}Q_{3}-J^2\right)} {8r_{*}J}.
	\nonumber
\end{IEEEeqnarray}
Note that the impact parameter $b_{\phi}$ is not fixed by these equations. In  fact it forms the extra parameter such that the edge of the shadow is a 2D surface and the shadow itself is a 3D volume as it should by in 5D. 

To get the shadow's edge as it appears for an observer we need to find what this translates to in camera coordinates. To do this we first write down the equations for the tangent vector components equations (obtained from \eqref{eq:velocities} by coordinate redefinition \eqref{eq:globalcoords}):
\begin{IEEEeqnarray}{rCl}
	v_{r} &=& \frac{\partial r}{\partial x_{i}}v_{x_{i}} = -\frac{d}{2}v_{x} ,\nonumber\\
	v_{\theta} &=&  \frac{\partial \theta}{\partial x_{i}}v_{x_{i}} = \frac{2}{d}v_{z},\nonumber\\
	v_{\phi} &=& \frac{\partial \phi}{\partial x_{i}}v_{x_{i}} = -\frac{v_{y}}{d\sin i} + \frac{v_{w}}{d\cos i}, \nonumber\\
	v_{\psi} &=& \frac{\partial \psi}{\partial x_{i}}v_{x_{i}} = -2\frac{v_{w}}{d\cos i}. \label{eq:compf}
\end{IEEEeqnarray}
We invert those equations to get the camera coordinate $\left(a,b,c\right)$ in terms of the tangent vector components $\left(v_{r},v_{\theta},v_{\phi},v_{\psi}\right)$ at the observer:
\begin{IEEEeqnarray}{rCl}
	a &=& \frac{1}{2}\left(\frac{d^2\sin i}{2 \tan\left(\frac{\alpha_{a}}{2}\right)}\left(\frac{v_{\phi}}{v_{r}} + \frac{1}{2}\frac{v_{\psi}}{v_{r}} \right) + 1\right),\\
	b &=& \frac{1}{2}\left(-\frac{d^2}{4 \tan\left(\frac{\alpha_{b}}{2}\right)}\frac{v_{\theta}}{v_{r}} + 1\right),\\
	c &=& \frac{1}{2}\left(\frac{d^2\cos i}{4 \tan\left(\frac{\alpha_{c}}{2}\right)}\frac{v_{\psi}}{v_{r}} + 1\right).
\end{IEEEeqnarray}
We can obtain the ratios of tangent vector components by using the first order equations of motion in the limit of $d \rightarrow \infty$:
\begin{IEEEeqnarray}{rCcCl}
	\frac{v_{\theta}}{v_{r}} &=& \left.  \frac{\dif \theta/\dif \tau}{\dif r/\dif \tau} \right\vert_{d \rightarrow \infty} &=&  \pm\frac{4}{E d^3}\sqrt{4K_{c} -\frac{L_{\phi}^2}{\sin^2(i)} - \frac{\left(2L_{\psi} - L_{\phi}\right)^2}{\cos^2(i)}} ,\\
	\frac{v_{\phi}}{v_{r}} &=& \left.  \frac{\dif \phi/\dif \tau}{\dif r/\dif \tau} \right\vert_{d \rightarrow \infty} 
	 &=& \frac{2}{Ed^3 \sin^2(i)\cos^2(i)} \left(L_{\phi} - 2\sin^2(i)L_{\psi}\right), \\
	\frac{v_{\psi}}{v_{r}} &=&  \left.  \frac{\dif \psi/\dif \tau}{\dif r/\dif \tau} \right\vert_{d \rightarrow \infty}
	&=&  \frac{4}{Ed^3 \cos^2(i)} \left(2L_{\psi} - L_{\phi}\right).
\end{IEEEeqnarray}
This gives us the final result:
\begin{IEEEeqnarray}{rCl}
	a(r_{*},b_{\phi}) &=& \frac{1}{2}\left(\frac{b_{\phi}}{d \tan\left(\frac{\alpha_{a}}{2}\right) \sin(i)} + 1\right),\\
	b(r_{*},b_{\phi}) &=& \frac{1}{2}\left(\frac{\pm 1}{d \tan\left(\frac{\alpha_{b}}{2}\right)}\sqrt{4\tilde{K_{c}}(r_{*}) -\frac{b_{\phi}^2}{\sin^2(i)} - \frac{\left(2b_{\psi}(r_{*}) - b_{\phi}\right)^2}{\cos^2(i)}} + 1\right),\\
	c(r_{*},b_{\phi}) &=& \frac{1}{2}\left(\frac{2b_{\psi}(r_{*}) - b_{\phi}}{d \tan\left(\frac{\alpha_{c}}{2}\right) \cos(i)} + 1\right).
\end{IEEEeqnarray}
The edge of the shadow is then the 2D surface in $\left(a,b,c\right)$ space that is parameterized by $b_{\phi}\in \mathbb{R}, r_{*} > 0$ within the range $0 \leq a,b,c \leq 1$.

\subsubsection{Non-rotating Black Hole \texorpdfstring{($J= 0$)}{}}
We find a double root for $R(r_*) =0$ when
\begin{IEEEeqnarray}{rCl}
	\tilde{K}_{c,0} &=& \frac{(4 r_{*} + Q_1)(4 r_{*} + Q_2) (4 r_{*} + Q_3)}{64 r_{*}^2},
\end{IEEEeqnarray}
where $r_{*}$ is the positive root of the polynomial:
\begin{IEEEeqnarray}{rCl}
	32 r_{*}^3 - 2 r_{*} \left(Q_{1}Q_{2} + Q_{1}Q_{3} + Q_{2}Q_{3}\right) - Q_{1}Q_{2}Q_{3}.
\end{IEEEeqnarray}
Again translating this to camera coordinates yields:
\begin{IEEEeqnarray}{rCl}
	a(b_{\phi},b_{\psi}) &=& \frac{1}{2}\left(\frac{b_{\phi}}{d \tan\left(\frac{\alpha_{a}}{2}\right) \sin(i)} + 1\right),\\
	b(b_{\phi},b_{\psi}) &=& \frac{1}{2}\left(\frac{\pm 1}{d\tan\left(\frac{\alpha_{b}}{2}\right)}\sqrt{4\tilde{K}_{c,0} -\frac{b_{\phi}^2}{\sin^2(i)} - \frac{\left(2b_{\psi} - b_{\phi}\right)^2}{\cos^2(i)}} + 1\right),\\
	c(b_{\phi},b_{\psi}) &=& \frac{1}{2}\left(\frac{2b_{\psi} - b_{\phi}}{d \tan\left(\frac{\alpha_{c}}{2}\right) \cos(i)} + 1\right).
\end{IEEEeqnarray}
Now $\tilde{K}_{c}$ is fixed at a certain value and the 2D shadow surface is instead parametrized  $b_{\phi}, L_\psi \in \mathbb{R}$ within the range $0 \leq a,b,c \leq 1$.

\bibliography{shadows}

\providecommand{\href}[2]{#2}\begingroup\raggedright\begin{thebibliography}{10}

\bibitem{Emparan:2008eg}
R.~Emparan and H.~S. Reall, \emph{{Black Holes in Higher Dimensions}},
  \href{https://doi.org/10.12942/lrr-2008-6}{\emph{Living Rev. Rel.} {\bfseries
  11} (2008) 6} [\href{https://arxiv.org/abs/0801.3471}{{\ttfamily
  0801.3471}}].

\bibitem{Herdeiro:2015aa}
C.~A.~R. Herdeiro and E.~Radu, \emph{Asymptotically flat black holes with
  scalar hair: a review},  \href{https://arxiv.org/abs/1504.08209}{{\ttfamily
  1504.08209}}.

\bibitem{Volkov:2016aa}
M.~S. Volkov, \emph{Hairy black holes in the xx-th and xxi-st centuries},
  \href{https://arxiv.org/abs/1601.08230}{{\ttfamily 1601.08230}}.

\bibitem{Emparan:2001wn}
R.~Emparan and H.~S. Reall, \emph{{A Rotating black ring solution in
  five-dimensions}},
  \href{https://doi.org/10.1103/PhysRevLett.88.101101}{\emph{Phys. Rev. Lett.}
  {\bfseries 88} (2002) 101101}
  [\href{https://arxiv.org/abs/hep-th/0110260}{{\ttfamily hep-th/0110260}}].

\bibitem{Grunau:2012ai}
S.~Grunau, V.~Kagramanova, J.~Kunz and C.~Lammerzahl, \emph{{Geodesic Motion in
  the Singly Spinning Black Ring Spacetime}},
  \href{https://doi.org/10.1103/PhysRevD.86.104002}{\emph{Phys. Rev.}
  {\bfseries D86} (2012) 104002}
  [\href{https://arxiv.org/abs/1208.2548}{{\ttfamily 1208.2548}}].

\bibitem{Gutowski:2004yv}
J.~B. Gutowski and H.~S. Reall, \emph{{General supersymmetric AdS(5) black
  holes}}, \href{https://doi.org/10.1088/1126-6708/2004/04/048}{\emph{JHEP}
  {\bfseries 04} (2004) 048}
  [\href{https://arxiv.org/abs/hep-th/0401129}{{\ttfamily hep-th/0401129}}].

\bibitem{Elvang:2004ds}
H.~Elvang, R.~Emparan, D.~Mateos and H.~S. Reall, \emph{{Supersymmetric black
  rings and three-charge supertubes}},
  \href{https://doi.org/10.1103/PhysRevD.71.024033}{\emph{Phys. Rev.}
  {\bfseries D71} (2005) 024033}
  [\href{https://arxiv.org/abs/hep-th/0408120}{{\ttfamily hep-th/0408120}}].

\bibitem{Gauntlett:2004qy}
J.~P. Gauntlett and J.~B. Gutowski, \emph{{General concentric black rings}},
  \href{https://doi.org/10.1103/PhysRevD.71.045002}{\emph{Phys. Rev.}
  {\bfseries D71} (2005) 045002}
  [\href{https://arxiv.org/abs/hep-th/0408122}{{\ttfamily hep-th/0408122}}].

\bibitem{Bena:2004de}
I.~Bena and N.~P. Warner, \emph{{One ring to rule them all ... and in the
  darkness bind them?}},
  \href{https://doi.org/10.4310/ATMP.2005.v9.n5.a1}{\emph{Adv. Theor. Math.
  Phys.} {\bfseries 9} (2005) 667}
  [\href{https://arxiv.org/abs/hep-th/0408106}{{\ttfamily hep-th/0408106}}].

\bibitem{Bena:2007kg}
I.~Bena and N.~P. Warner, \emph{{Black holes, black rings and their
  microstates}}, \href{https://doi.org/10.1007/978-3-540-79523-0_1}{\emph{Lect.
  Notes Phys.} {\bfseries 755} (2008) 1}
  [\href{https://arxiv.org/abs/hep-th/0701216}{{\ttfamily hep-th/0701216}}].

\bibitem{Breckenridge:1996is}
J.~C. Breckenridge, R.~C. Myers, A.~W. Peet and C.~Vafa, \emph{{D-branes and
  spinning black holes}},
  \href{https://doi.org/10.1016/S0370-2693(96)01460-8}{\emph{Phys. Lett.}
  {\bfseries B391} (1997) 93}
  [\href{https://arxiv.org/abs/hep-th/9602065}{{\ttfamily hep-th/9602065}}].

\bibitem{Elvang:2004rt}
H.~Elvang, R.~Emparan, D.~Mateos and H.~S. Reall, \emph{{A Supersymmetric black
  ring}}, \href{https://doi.org/10.1103/PhysRevLett.93.211302}{\emph{Phys. Rev.
  Lett.} {\bfseries 93} (2004) 211302}
  [\href{https://arxiv.org/abs/hep-th/0407065}{{\ttfamily hep-th/0407065}}].

\bibitem{Papnoi:2014aaa}
U.~Papnoi, F.~Atamurotov, S.~G. Ghosh and B.~Ahmedov, \emph{{Shadow of
  five-dimensional rotating Myers-Perry black hole}},
  \href{https://doi.org/10.1103/PhysRevD.90.024073}{\emph{Phys. Rev.}
  {\bfseries D90} (2014) 024073}
  [\href{https://arxiv.org/abs/1407.0834}{{\ttfamily 1407.0834}}].

\bibitem{Nitta:2011in}
D.~Nitta, T.~Chiba and N.~Sugiyama, \emph{{Shadows of Colliding Black Holes}},
  \href{https://doi.org/10.1103/PhysRevD.84.063008}{\emph{Phys. Rev.}
  {\bfseries D84} (2011) 063008}
  [\href{https://arxiv.org/abs/1106.2425}{{\ttfamily 1106.2425}}].

\bibitem{Yumoto:2012kz}
A.~Yumoto, D.~Nitta, T.~Chiba and N.~Sugiyama, \emph{{Shadows of Multi-Black
  Holes: Analytic Exploration}},
  \href{https://doi.org/10.1103/PhysRevD.86.103001}{\emph{Phys. Rev.}
  {\bfseries D86} (2012) 103001}
  [\href{https://arxiv.org/abs/1208.0635}{{\ttfamily 1208.0635}}].

\bibitem{Bohn:2014xxa}
A.~Bohn, W.~Throwe, F.~H{\'e}bert, K.~Henriksson, D.~Bunandar, M.~A. Scheel
  et~al., \emph{{What does a binary black hole merger look like?}},
  \href{https://doi.org/10.1088/0264-9381/32/6/065002}{\emph{Class. Quant.
  Grav.} {\bfseries 32} (2015) 065002}
  [\href{https://arxiv.org/abs/1410.7775}{{\ttfamily 1410.7775}}].

\bibitem{Patil:2016oav}
M.~Patil, P.~Mishra and D.~Narasimha, \emph{{Curious case of gravitational
  lensing by binary black holes: a tale of two photon spheres, new relativistic
  images and caustics}},
  \href{https://doi.org/10.1103/PhysRevD.95.024026}{\emph{Phys. Rev.}
  {\bfseries D95} (2017) 024026}
  [\href{https://arxiv.org/abs/1610.04863}{{\ttfamily 1610.04863}}].

\bibitem{Shipley:2016omi}
J.~Shipley and S.~R. Dolan, \emph{{Binary black hole shadows, chaotic
  scattering and the Cantor set}},
  \href{https://doi.org/10.1088/0264-9381/33/17/175001}{\emph{Class. Quant.
  Grav.} {\bfseries 33} (2016) 175001}
  [\href{https://arxiv.org/abs/1603.04469}{{\ttfamily 1603.04469}}].

\bibitem{Wang:2017qhh}
M.~Wang, S.~Chen and J.~Jing, \emph{{Shadows of Bonnor black dihole by chaotic
  lensing}}, \href{https://doi.org/10.1103/PhysRevD.97.064029}{\emph{Phys.
  Rev.} {\bfseries D97} (2018) 064029}
  [\href{https://arxiv.org/abs/1710.07172}{{\ttfamily 1710.07172}}].

\bibitem{Cunha:2018gql}
P.~V.~P. Cunha, C.~A.~R. Herdeiro and M.~J. Rodriguez, \emph{{Does the black
  hole shadow probe the event horizon geometry?}},
  \href{https://doi.org/10.1103/PhysRevD.97.084020}{\emph{Phys. Rev.}
  {\bfseries D97} (2018) 084020}
  [\href{https://arxiv.org/abs/1802.02675}{{\ttfamily 1802.02675}}].

\bibitem{Cunha:2018cof}
P.~V.~P. Cunha, C.~A.~R. Herdeiro and M.~J. Rodriguez, \emph{{Shadows of Exact
  Binary Black Holes}},
  \href{https://doi.org/10.1103/PhysRevD.98.044053}{\emph{Phys. Rev.}
  {\bfseries D98} (2018) 044053}
  [\href{https://arxiv.org/abs/1805.03798}{{\ttfamily 1805.03798}}].

\bibitem{Cunha:2016bjh}
P.~V.~P. Cunha, J.~Grover, C.~Herdeiro, E.~Radu, H.~Runarsson and A.~Wittig,
  \emph{{Chaotic lensing around boson stars and Kerr black holes with scalar
  hair}}, \href{https://doi.org/10.1103/PhysRevD.94.104023}{\emph{Phys. Rev.}
  {\bfseries D94} (2016) 104023}
  [\href{https://arxiv.org/abs/1609.01340}{{\ttfamily 1609.01340}}].

\bibitem{Grunau:2012ri}
S.~Grunau, V.~Kagramanova and J.~Kunz, \emph{{Geodesic Motion in the (Charged)
  Doubly Spinning Black Ring Spacetime}},
  \href{https://doi.org/10.1103/PhysRevD.87.044054}{\emph{Phys. Rev.}
  {\bfseries D87} (2013) 044054}
  [\href{https://arxiv.org/abs/1212.0416}{{\ttfamily 1212.0416}}].

\bibitem{Cunha:2015yba}
P.~V.~P. Cunha, C.~A.~R. Herdeiro, E.~Radu and H.~F. Runarsson, \emph{{Shadows
  of Kerr black holes with scalar hair}},
  \href{https://doi.org/10.1103/PhysRevLett.115.211102}{\emph{Phys. Rev. Lett.}
  {\bfseries 115} (2015) 211102}
  [\href{https://arxiv.org/abs/1509.00021}{{\ttfamily 1509.00021}}].

\bibitem{Cunha:2018acu}
P.~V.~P. Cunha and C.~A.~R. Herdeiro, \emph{{Shadows and strong gravitational
  lensing: a brief review}},
  \href{https://doi.org/10.1007/s10714-018-2361-9}{\emph{Gen. Rel. Grav.}
  {\bfseries 50} (2018) 42} [\href{https://arxiv.org/abs/1801.00860}{{\ttfamily
  1801.00860}}].

\bibitem{Hawking:1976jb}
S.~W. Hawking, \emph{{Gravitational Instantons}},
  \href{https://doi.org/10.1016/0375-9601(77)90386-3}{\emph{Phys. Lett.}
  {\bfseries A60} (1977) 81}.

\bibitem{Gibbons:1979zt}
G.~W. Gibbons and S.~W. Hawking, \emph{{Gravitational Multi - Instantons}},
  \href{https://doi.org/10.1016/0370-2693(78)90478-1}{\emph{Phys. Lett.}
  {\bfseries 78B} (1978) 430}.

\bibitem{Gibbons:1999uv}
G.~W. Gibbons and C.~A.~R. Herdeiro, \emph{{Supersymmetric rotating black holes
  and causality violation}},
  \href{https://doi.org/10.1088/0264-9381/16/11/311}{\emph{Class. Quant. Grav.}
  {\bfseries 16} (1999) 3619}
  [\href{https://arxiv.org/abs/hep-th/9906098}{{\ttfamily hep-th/9906098}}].

\bibitem{Herdeiro:2000ap}
C.~A.~R. Herdeiro, \emph{{Special properties of five-dimensional BPS rotating
  black holes}},
  \href{https://doi.org/10.1016/S0550-3213(00)00335-7}{\emph{Nucl. Phys.}
  {\bfseries B582} (2000) 363}
  [\href{https://arxiv.org/abs/hep-th/0003063}{{\ttfamily hep-th/0003063}}].

\bibitem{Herdeiro:2002ft}
C.~A.~R. Herdeiro, \emph{{Spinning deformations of the D1 - D5 system and a
  geometric resolution of closed timelike curves}},
  \href{https://doi.org/10.1016/S0550-3213(03)00484-X}{\emph{Nucl. Phys.}
  {\bfseries B665} (2003) 189}
  [\href{https://arxiv.org/abs/hep-th/0212002}{{\ttfamily hep-th/0212002}}].

\bibitem{Drukker:2004zm}
N.~Drukker, \emph{{Supertube domain walls and elimination of closed time-like
  curves in string theory}},
  \href{https://doi.org/10.1103/PhysRevD.70.084031}{\emph{Phys. Rev.}
  {\bfseries D70} (2004) 084031}
  [\href{https://arxiv.org/abs/hep-th/0404239}{{\ttfamily hep-th/0404239}}].

\bibitem{Gauntlett:1998fz}
J.~P. Gauntlett, R.~C. Myers and P.~K. Townsend, \emph{{Black holes of D = 5
  supergravity}},
  \href{https://doi.org/10.1088/0264-9381/16/1/001}{\emph{Class. Quant. Grav.}
  {\bfseries 16} (1999) 1}
  [\href{https://arxiv.org/abs/hep-th/9810204}{{\ttfamily hep-th/9810204}}].

\bibitem{Diemer:2013fza}
V.~Diemer and J.~Kunz, \emph{{Supersymmetric rotating black hole spacetime
  tested by geodesics}},
  \href{https://doi.org/10.1103/PhysRevD.89.084001}{\emph{Phys. Rev.}
  {\bfseries D89} (2014) 084001}
  [\href{https://arxiv.org/abs/1312.6540}{{\ttfamily 1312.6540}}].

\bibitem{Grover:2017mhm}
J.~Grover and A.~Wittig, \emph{{Black Hole Shadows and Invariant Phase Space
  Structures}}, \href{https://doi.org/10.1103/PhysRevD.96.024045}{\emph{Phys.
  Rev.} {\bfseries D96} (2017) 024045}
  [\href{https://arxiv.org/abs/1705.07061}{{\ttfamily 1705.07061}}].

\bibitem{Wang:2018aa}
M.~Wang, S.~Chen and J.~Jing, \emph{Chaotic shadow of a non-kerr rotating
  compact object with quadrupole mass moment},
  \href{https://arxiv.org/abs/1801.02118}{{\ttfamily 1801.02118}}.

\bibitem{Hertog:inprep1}
T.~Hertog, T.~Lemmens and B.~Vercnocke, \emph{{The Face of Fuzzballs}},
  {\emph{in preparation} }.

\bibitem{Tyukov:2017uig}
A.~Tyukov, R.~Walker and N.~P. Warner, \emph{{Tidal Stresses and Energy Gaps in
  Microstate Geometries}},
  \href{https://doi.org/10.1007/JHEP02(2018)122}{\emph{JHEP} {\bfseries 02}
  (2018) 122} [\href{https://arxiv.org/abs/1710.09006}{{\ttfamily
  1710.09006}}].

\bibitem{Bena:2018mpb}
I.~Bena, E.~J. Martinec, R.~Walker and N.~P. Warner, \emph{{Early Scrambling
  and Capped BTZ Geometries}},
  \href{https://arxiv.org/abs/1812.05110}{{\ttfamily 1812.05110}}.

\bibitem{Giddings:2016btb}
S.~B. Giddings and D.~Psaltis, \emph{{Event Horizon Telescope Observations as
  Probes for Quantum Structure of Astrophysical Black Holes}},
  \href{https://doi.org/10.1103/PhysRevD.97.084035}{\emph{Phys. Rev.}
  {\bfseries D97} (2018) 084035}
  [\href{https://arxiv.org/abs/1606.07814}{{\ttfamily 1606.07814}}].

\bibitem{Gauntlett:2002nw}
J.~P. Gauntlett, J.~B. Gutowski, C.~M. Hull, S.~Pakis and H.~S. Reall,
  \emph{{All supersymmetric solutions of minimal supergravity in five-
  dimensions}}, \href{https://doi.org/10.1088/0264-9381/20/21/005}{\emph{Class.
  Quant. Grav.} {\bfseries 20} (2003) 4587}
  [\href{https://arxiv.org/abs/hep-th/0209114}{{\ttfamily hep-th/0209114}}].

\bibitem{Gauntlett:2003fk}
J.~P. Gauntlett and J.~B. Gutowski, \emph{{All supersymmetric solutions of
  minimal gauged supergravity in five-dimensions}},
  \href{https://doi.org/10.1103/PhysRevD.70.089901,
  10.1103/PhysRevD.68.105009}{\emph{Phys. Rev.} {\bfseries D68} (2003) 105009}
  [\href{https://arxiv.org/abs/hep-th/0304064}{{\ttfamily hep-th/0304064}}].

\bibitem{Denef:2000nb}
F.~Denef, \emph{{Supergravity flows and D-brane stability}},
  \href{https://doi.org/10.1088/1126-6708/2000/08/050}{\emph{JHEP} {\bfseries
  08} (2000) 050} [\href{https://arxiv.org/abs/hep-th/0005049}{{\ttfamily
  hep-th/0005049}}].

\bibitem{Bates:2003vx}
B.~Bates and F.~Denef, \emph{{Exact solutions for supersymmetric stationary
  black hole composites}},
  \href{https://doi.org/10.1007/JHEP11(2011)127}{\emph{JHEP} {\bfseries 11}
  (2011) 127} [\href{https://arxiv.org/abs/hep-th/0304094}{{\ttfamily
  hep-th/0304094}}].

\bibitem{Gibbons:2013tqa}
G.~W. Gibbons and N.~P. Warner, \emph{{Global structure of five-dimensional
  fuzzballs}},
  \href{https://doi.org/10.1088/0264-9381/31/2/025016}{\emph{Class. Quant.
  Grav.} {\bfseries 31} (2014) 025016}
  [\href{https://arxiv.org/abs/1305.0957}{{\ttfamily 1305.0957}}].

\bibitem{Breckenridge:1996sn}
J.~C. Breckenridge, D.~A. Lowe, R.~C. Myers, A.~W. Peet, A.~Strominger and
  C.~Vafa, \emph{{Macroscopic and microscopic entropy of near extremal spinning
  black holes}},
  \href{https://doi.org/10.1016/0370-2693(96)00553-9}{\emph{Phys. Lett.}
  {\bfseries B381} (1996) 423}
  [\href{https://arxiv.org/abs/hep-th/9603078}{{\ttfamily hep-th/9603078}}].

\bibitem{Bena:2005ni}
I.~Bena, P.~Kraus and N.~P. Warner, \emph{{Black rings in Taub-NUT}},
  \href{https://doi.org/10.1103/PhysRevD.72.084019}{\emph{Phys. Rev.}
  {\bfseries D72} (2005) 084019}
  [\href{https://arxiv.org/abs/hep-th/0504142}{{\ttfamily hep-th/0504142}}].

\bibitem{Psaltis:2010ww}
D.~Psaltis and T.~Johannsen, \emph{{A Ray-Tracing Algorithm for Spinning
  Compact Object Spacetimes with Arbitrary Quadrupole Moments. I. Quasi-Kerr
  Black Holes}},
  \href{https://doi.org/10.1088/0004-637X/745/1/1}{\emph{Astrophys. J.}
  {\bfseries 745} (2012) 1} [\href{https://arxiv.org/abs/1011.4078}{{\ttfamily
  1011.4078}}].

\end{thebibliography}\endgroup
\bibliographystyle{JHEP}

\end{document}